\documentclass[letterpaper]{article}
\pdfoutput=1
\usepackage[T1]{fontenc}
\usepackage{geometry}
\usepackage{setspace}
\usepackage[super,sort&compress]{natbib}
\usepackage{graphicx}
\usepackage{float}
\newfloat{scheme}{htbp}{los}
\floatname{scheme}{Scheme}
\usepackage{amssymb}
\usepackage{amsmath}
\usepackage{dsfont}
\usepackage{commath}
\usepackage{bm}
\usepackage{braket}
\usepackage{hyperref}
\usepackage{siunitx}
\usepackage{xcolor}
\usepackage[capitalize]{cleveref}
\newcommand{\be}{\begin{equation}}
	\newcommand{\ee}{\end{equation}}

\newcommand{\eq}[1]{Eq.~\eqref{#1}}

\renewcommand{\bf}[1]{\mathbf{#1}}

\usepackage{authblk}
\author[1]{Ruoxi Liu}
\author[1]{Xiaotong Zhu}
\author[1]{Bing Gu*}
\affil[1]{Department of Chemistry and Department of Physics, Westlake University, Hangzhou, Zhejiang 310030, China}

\title{Crossing Seam Blockade}
\date{*Email: gubing@westlake.edu.cn}

\begin{document}
	
	\maketitle
	
	\begin{abstract}
		Electronic degeneracies and near-degeneracies including conical intersections and avoided crossings, typically accompanied by strong vibronic couplings and nonadiabatic transitions, play fundamental roles in photochemical, photophysical and photobiological processes. 
		However, its implications on excited-state chemical reactivities are not fully understood. 
		In this theoretical study, we report a surprising phenomena that an open reaction channel can be \emph{completely} blocked by a crossing seam in the molecular configuration space. Specifically,  by numerically exact ab initio nonadiabatic full quantum geometrical molecular dynamics simulations, we show that the singlet fission channel in the hydrogen chain H$_4$, previously identified as a minimal model for singlet fission, is blocked due to electronic quantum geometry. We provide a chemically intuitive picture to understand this effect. Our results not only reveal a new mechanism for controlling photochemical reactions, but may also elucidate the mechanism of singlet fission.
	\end{abstract}

	\section{Introduction}
	
	Virtually all ultrafast photochemical, photophysical, and photobiological processes are dictated by regions of electronic state degeneracy and near-degeneracy in the configuration space,  including avoided crossings and conical intersections (CIs) \cite{domckeConicalIntersectionsTheory2011,baerAdvancesChemicalPhysics2002,nakamuraNonadiabaticTransitionConcepts2012,domckeRoleConicalIntersections2012} in molecular configuration space. 
	Around these regions, the Born-Oppenheimer approximation \cite{bornZurQuantentheorieMolekeln1927} breaks down and nonadiabatic electronic transitions occur in femtosecond timescales due to strong electron-nuclear (vibronic) couplings\cite{tannorIntroductionQuantumMechanics2007}. 
	Such ultrafast reactions  play vital roles in physical, chemical, and biological processes, including rhodopsin photoisomerization as a primary event of vision \cite{kandoriPhotoisomerizationRhodopsin2001,polliConicalIntersectionDynamics2010}, 
	excited-state deactivation of DNA nucleobases via CIs that provides intrinsic photoprotection against ultraviolet radiation \cite{barbattiRelaxationMechanismsUVphotoexcited2010,prokhorenkoNewInsightsPhotophysics2016}, 
	and ozone photodissociation in the Chappuis band \cite{grebenshchikovNewTheoreticalInvestigations2007}. 
	
	The importance of exact and quasi electronic degeneracy, particularly the CI seam, in nonradiative electronic relaxation (i.e., internal conversion) has been well-recognized in many molecules \cite{yarkonyIntersectingConicalIntersection2001,fazziUltrafastInternalConversion2012,xieNonadiabaticTunnelingPhotodissociation2016,yangConicalIntersectionsParticle2016,yuePerformanceTDDFTSpinflip2018,guanEnablingUnifiedDescription2021,huInitioMolecularCavity2023,pollakEffectOpticalCavity2025,jiangSuperresolutionFemtosecondElectron2025,zhangUnderstandingMolecularMotions2025,dustonConicalIntersectionsElectronic2025a}. In the language of conventional Born-Huang framework, this is because the nonadiabatic coupling is significant around the CI. Beyond providing a funnel for internal conversion, the CIs also introduce the geometric phase effect\cite{berryQuantalPhaseFactors1984,meadGeometricPhaseMolecular1992,ryabinkinGeometricPhaseEffects2013a,ryabinkinGeometricPhaseEffects2017,faragPolaritonInducedConical2021}, whereby electronic wavefunctions acquire a phase change when encircling a CI. So does the nuclear wavefunction to make the total wavefunction single-valued. In chemical reaction dynamics,  this geometric phase is typically manifested as destructive nuclear quantum interference. Furthermore, there are diagonal Born-Oppenheimer corrections \cite{handyDiagonalCorrectionBorn1986}, a scalar potential that alters the the adiabatic potential energy surfaces (APES), and second-order nonadiabatic couplings \cite{baerBornOppenheimerConical2006}, that can also induce electronic transitions.
	Understanding all such non-Born-Oppenheimer effects is challenging within the conventional Born-Huang framework\cite{bornDynamicalTheoryCrystal1955}, as all vibronic couplings diverge at CIs. Specifically, both the first- and second-derivative couplings become singular at points of degeneracy; the diagonal Born-Oppenheimer corrections diverge at CIs and are not even integrable\cite{matselyukhDecoherenceRevivalAttosecond2022}; the vector potential associated with the geometric phase carries a gauge-dependent singular branch cut. Such divergences make molecular modeling problematic, often requiring approximate treatments such as quasi-diabatization\cite{nakamuraDirectCalculationDiabatic2001,subotnikConstructingDiabaticStates2008} and vibronic coupling model Hamiltonians\cite{kouppelMultimodeMolecularDynamics1984,koppelDiabaticRepresentationMethods2004}. 
	
	We recently proposed a quantum geometrical molecular dynamics framework that unifies all non-Born-Oppenheimer effects into a single global electronic overlap matrix \cite{guDiscreteVariableLocalDiabatic2023,guNonadiabaticConicalIntersection2024b,zhuMakingPeaceRandom2024a,xieQuantumGeometricalMolecular2025}, thus eliminating the divergences associated with derivative couplings and vector potential singularities. This framework not only provides a geometric picture for understanding photochemical reactions but also a computational framework for exact ab initio simulations in regions of electronic degeneracy and near-degeneracy\cite{shaExponentialConvergenceLocal2026}. 
	
	In this work, by numerically exact full quantum dynamics simulations, we unveil a surprising effect, that is, an open reaction channel can be completely blocked by a crossing seam in the configuration space. This effect is referred to as crossing seam blockade (CSB). This phenomenon is demonstrated in a singlet fission (SF)\cite{popeElectronicProcessesOrganic1999,smithSingletFission2010,smithRecentAdvancesSinglet2013} process using the hydrogen chain. 
	The H$_4$ chain was identified as a model molecule for understanding SF\cite{minamiDiradicalCharacterView2012,nakanoFullConfigurationInteraction2012,nakanoOpenShellCharacterBasedMolecularDesign2017}, which is an ultrafast photochemical process in which one photoexcited singlet exciton converts into two spin-correlated triplet excitons\cite{popeElectronicProcessesOrganic1999}. The SF process is feasible in this molecule according to electronic structure analysis \cite{thalmannRoleRadicalCharacter2024,claudinoModelingSingletFission2023a}. H$_4$ contains multi-excitation configurations makes it an ideal molecule for studying SF.
	
	The exact nonadiabatic molecular quantum dynamic simulations is made possible by combining full configuration interaction (FCI)  electronic structure method and the geometric quantum dynamics. A local spin analysis is employed  for characterizing the low-lying electronic states, thereby revealing the singlet fission channel.
	There are two apparent mechanism for the blockade: the first due to the barrier on the (adiabatic) potential energy surface, and the second due to the geometric phase effect. 
	By increasing the initial kinetic energy along the reaction coordinate way beyond the barrier height, the blockade still persists, thus  ruling out the barrier as the mechanism. 
	Moreover, by turning off the geometric phase in the quantum dynamics simulation, we find that the geometric phase,  while does influence the nonadiabatic dynamics in the long time, does not fully account for the blockade.
	We finally proposed that the blockade is completely due to the crossing seam, and provide an intuitive geometric picture to explain the CSB effect, showing that the crossing seam  is associated with abrupt changes of electronic state character across the entire seam.
	
	We further examine if the quantum nature of nuclei is essential for the CSB phenomena to occur. To do so, we carry out the Ehrenfest dynamics \cite{berneClassicalQuantumDynamics1998,tullyEhrenfestDynamicsQuantum2023}, a widely used mixed quantum-classical method that treats nuclear motion classically. In the Ehrenfest dynamics, the singlet fission  does in fact occur, demonstrating that CSB is a nuclear quantum effect that cannot be captured without a full quantum treatment of electrons and nuclei. 
	
	This work not only demonstrates the utility of quantum geometrical dynamics to tackle molecules with complex quantum geometry of electronic states, but also reveals a new mechanism for controlling photochemical reactions through manipulation of the topology of crossing seam space, while reassessing the conditions for SF occurrence in the H$_4$ system from a dynamical perspective\cite{musserEvidenceConicalIntersection2015}. 
	
	\section{Results and Discussion}
	
	\subsection{Singlet Fission in H$_4$}
	
	We first reveal the singlet fission reaction channel in the H$_4$ chain system by electronic structure analysis. We label the four hydrogen atoms, sequentially H1--H4 along the molecular axis with a fixed terminal separation of $R_{\mathrm{H1-H4}}=11.3$ Bohr. Denoting the positions of the two inner atoms H2 and H3 along the molecular axis as $x_2$ and $x_3$, their motion is described by two collective reaction coordinates: the symmetric stretch mode $q_1 = \frac{1}{\sqrt{2}}(x_2 + x_3)$ and the antisymmetric stretch mode $q_2 = \frac{1}{\sqrt{2}}(x_2 - x_3)$. The APES of the four lowest singlet states (S$_0$--S$_3$) were constructed in the $(q_1, q_2)$ coordinate space in the level of FCI/cc-pVDZ\cite{dunningGaussianBasisSets1989b,kendallElectronAffinitiesFirstrow1992} (Fig.\ S1a). The excited state PESs exhibit clear near-degeneracy features, with a crossing seam between S$_1$ and S$_2$ that extends across the entire configure space (Fig.\ S1b).
	
\begin{figure*}[t]
		\centering
		\includegraphics[width=1.0\textwidth]{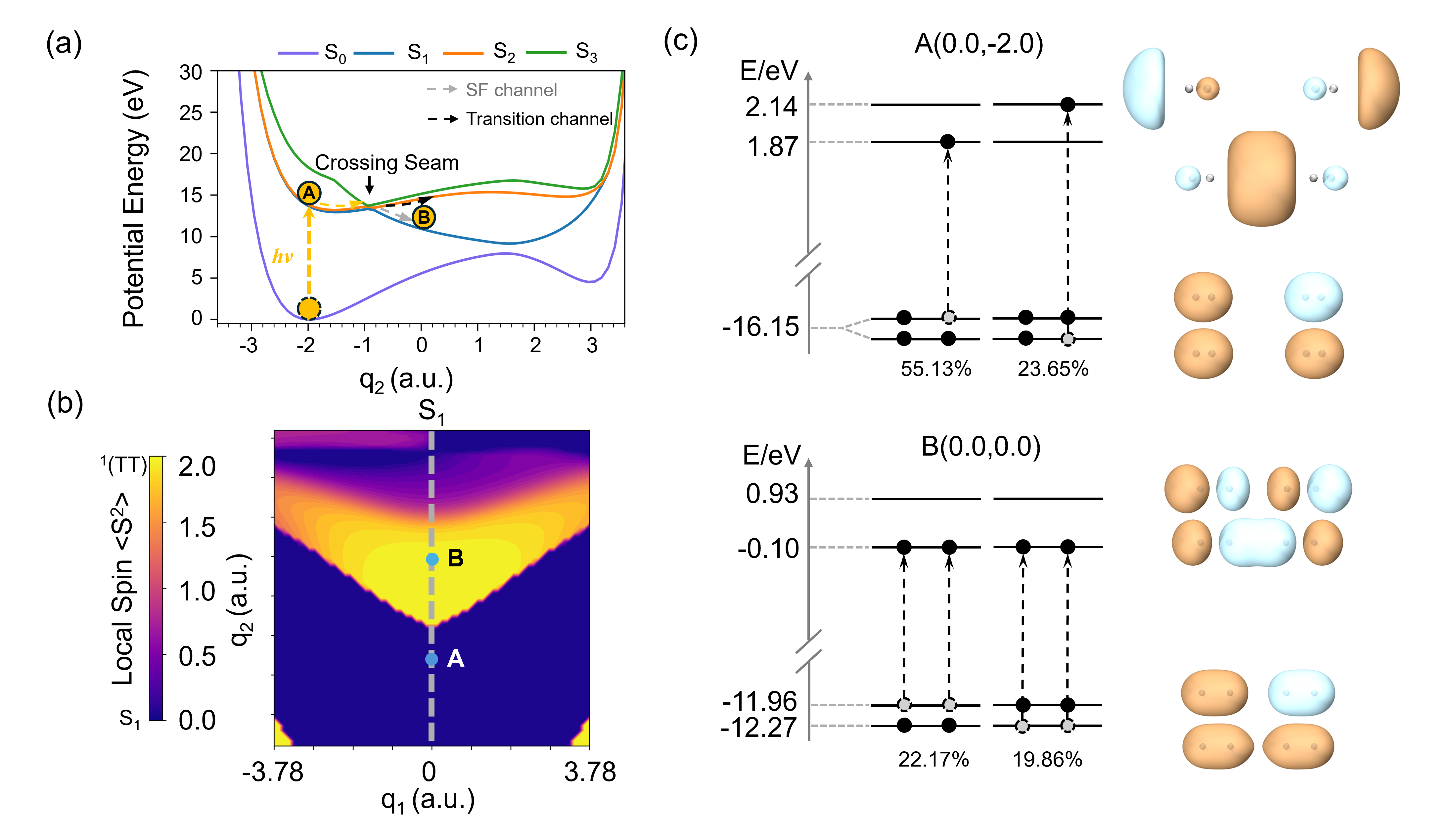}
		\caption{Electronic structure along the $q_1 = 0$ reaction coordinate and identification of reaction pathways.
			(a) Potential energy curves of the singlet states along $q_1 = 0$. Points~A and~B mark two representative geometries. Following photoexcitation from S$_0$ to S$_1$, two candidate pathways are indicated: a transition channel via the crossing seam in S$_1$ and S$_2$ as well as a SF channel along the S$_1$ surface toward the $^1(\mathrm{TT})$ region.
			(b) Fragment local spin map of the S$_1$ state, $S_{\mathrm{loc}}^{2}(F)$ with $F = \{\mathrm{H1}, \mathrm{H2}\}$, in the $(q_1, q_2)$ coordinate space. Regions where $S_{\mathrm{loc}}^{2}(F) \approx 2$ indicate triplet-pair character.
			(c) Dominant excitation configurations of the S$_1$ state at points~A and~B, together with the relevant molecular orbitals (HOMO$-1$, HOMO, LUMO, and LUMO$+1$). Orbital isosurfaces are rendered at $\mathrm{isosurface value} = 0.036$.}
		\label{fig:pathways}
	\end{figure*}

	A reaction pathway can be identified as the $q_1 = 0$ slice. The S$_0$ state possesses a global energy minimum at $q_2 \approx -2.0$~Bohr (point~A in \cref{fig:pathways}a), with electronic structure analogous to two nearly independent H$_2$ molecules ($R_{\mathrm{H1{-}H2}} \approx 1.42$~Bohr, $R_{\mathrm{H2{-}H3}} \approx 8.50$~Bohr). Moving along the positive $q_2$ direction, the molecule passes through point~B ($q_2 = 0.0$), where $R_{\mathrm{H1{-}H2}}$ is elongated to 2.84~Bohr and $R_{\mathrm{H2{-}H3}}$ is shortened to 5.67~Bohr. Continuing further in the $q_2 > 0$ direction: the intramolecular H--H bonds progressively rupture and the four hydrogen atoms become increasingly isolated. This process is accompanied by a dramatic increase in electronic structure complexity---despite the simplicity of the molecule,  electron correlation effects become significant in the hydrogen bond-breaking region, causing the Hartree-Fock method to qualitatively fail in describing even the ground state (Fig.\ S2). All four electrons have to be included in the active space; the CASSCF(4e,4o) energy is close to the FCI energy within the chosen basis set.
	
	Screening the APES with the energy criterion $E(\mathrm{S}_1) \ge 2E(\mathrm{T}_1)$ reveals that extensive geometric regions satisfy the energetic prerequisite for SF (Fig.\ S3), indicating the existence of a potential SF reaction channel in this system. To further identify the $^1(\mathrm{TT})$ state, we performed fragment local spin analysis on the S$_1$ state \cite{chauvinApplicationsTopologicalMethods2016,martinpendasLocalSpinOpen2021}. The local spin $S_{\mathrm{loc}}^{2}(F)$ distribution computed for fragment $F = \{\mathrm{H1}, \mathrm{H2}\}$ (\cref{fig:pathways}b) shows that there is a large region in the configuration space with $q_2 > 0$ with the local spin value approaching ~2 (triplet character) while the total spin of the system remains singlet. This is the hallmark signature of the triplet-pair coupled singlet state $^1(\mathrm{TT})$. 
	These regions exhibiting triplet character coincide well with those satisfying the SF energy criterion. Based on the electronic structure analysis, we can envision the following reaction process (\cref{fig:pathways}a): upon photoexcitation from S$_0$ to S$_1$, the system evolves along the S$_1$ surface toward the $^1(\mathrm{TT})$ region, with the SF reaction being energetically favorable.
	
	Wavefunction analysis of the S$_1$ state at two representative geometries (\cref{fig:pathways}c) reveals the qualitative changes in electronic structure along the SF channel. At point~A, the S$_1$ state is dominated by single-excitation configurations: the HOMO~$\rightarrow$~LUMO single excitation accounts for 55.13\% and the HOMO$-1$~$\rightarrow$~LUMO$+1$ single excitation for 23.65\%, exhibiting typical local excitation character. At point~B, however, the electronic structure undergoes a qualitative change: the dominant configurations are almost entirely of double-excitation character, including HOMO~$\rightarrow$~LUMO double excitation, HOMO$-1$~$\rightarrow$~LUMO double excitation, and similar configurations. This transition from single-excitation to double-excitation dominated electronic structure corroborates the local spin analysis, demonstrating that the potential SF region is characterized by electronic states with dominant double-excitation character, consistent with previous studies on pentacene and related molecules\cite{zirzlmeierSingletFissionPentacene2015,casanovaTheoreticalModelingSinglet2018}.

	The electronic structure analysis clearly demonstrates that the H$_4$ chain satisfies the requisite energetic and spin conditions for SF at the electronic structure level: the S$_1$ state exhibits pronounced triplet-pair local spin character in the product region, where the SF energy criterion is met as well. 
	It is therefore reasonable to expect that singlet fission would occur in this molecule\cite{claudinoModelingSingletFission2023a,roseiroEnvironmentalEffectsSinglet2022,thalmannRoleRadicalCharacter2024}.

	\subsection{Geometric Singlet Fission Quantum Dynamics}
	
	We employ geometric quantum molecular dynamics\cite{guDiscreteVariableLocalDiabatic2023,guNonadiabaticConicalIntersection2024b} to solve the time-dependent molecular Schr\"{o}dinger equation, i.e., 
	\be
	i \pd{ \bm \chi(t)}{t} = \int \dif \bf q' T_\text{N}(\bf q, \bf q') \bf A(\bf q, \bf q') \bm \chi(\bf q') + \bf V(\bf q) \bm \chi(\bf q, t)  
	\label{eq:geom}
	\ee 
	where $\bm \chi = \set{\chi_1(\bf q), \cdots, \chi_N(\bf q)}$ is a column vector of the nuclear wave packets on $N$ APESs, $T_\text{N}(\bf q, \bf q')$ is the coordinate representation of the nuclear kinetic energy operator, which is dressed by the global electronic overlap matrix 
	\be A_{\beta \alpha}(\bf q, \bf q') = \braket{\phi_\beta(\bf q) | \phi_\alpha(\bf q')}_{\bf r}, \ee $\bf V(\bf q)$ is a diagonal matrix of the APES.
	\cref{eq:geom} is exact in the sense that it does not involve further approximations apart from truncation of electronic states. 
	We employ the discrete variable representation (DVR)\cite{littlejohnGeneralFrameworkDiscrete2002,lightDiscreteVariableRepresentationsTheir2000} basis set to regularize \eq{eq:geom}, such that the adiabatic electronic states are calculated by quantum chemistry at the DVR grid points, 
	$
	H_{\mathrm{el}}(\mathbf{q}_{\mathbf{n}}) \vert \phi_\alpha(\mathbf{q}_{\mathbf{n}}) \rangle = V_\alpha(\mathbf{q}_{\mathbf{n}}) \vert \phi_\alpha(\mathbf{q}_{\mathbf{n}}) \rangle,
	$
	where $H_{\mathrm{el}} = H - \hat{T}_{\mathrm{N}}$ is the electronic Hamiltonian and $V_\alpha(\mathbf{q}_{\mathbf{n}})$ is the $\alpha$th adiabatic electronic energy at geometry $\mathbf{q}_{\mathbf{n}}$.
	One main practical advantage of the geometric approach is that no smoothness or phase consistency condition required in the Born-Huang framework is imposed on the electronic states $\phi_\alpha(\mathbf{q}_{\mathbf{n}})$ across the configuration space. This makes the geometric quantum dynamics particularly convenient for ab initio modeling as the many-electron states obtained from the electronic structure solver always carry random phases (i.e., random gauge fixing) \cite{zhuMakingPeaceRandom2024a}.  
	
	Conceptually, the geometric quantum dynamics provides a different picture  emphasizing the quantum geometry of electronic states from the Born-Huang picture understanding nuclear motion \cite{xieQuantumGeometricalMolecular2025}. There are no  derivative couplings appearing in the equation of motion; all non-Born-Oppenheimer effects  including nonadiabatic transitions and geometric phase effects are encoded in the  electronic overlap matrix. The Born-Oppenheimer dynamics correspond to the limit of trivial geometry with $A_{\beta \alpha} = \delta_{\beta \alpha}$. In this geometric picture, the nuclear motion is determined jointly by the topology of the potential energy surfaces and the quantum geometry of electronic states. Nonadiabatic transitions occur because the excited electronic state at one configuration is similar to the ground  state of another configuration. Qualitatively speaking, the nuclear motion is dominated by the APES in the region where the electronic character does not change too much, however, the electronic quantum geometry will take over when  the electronic states vary significantly.  This implies that the problem of electron correlation in quantum chemistry is intimately related to the problem of electron-nuclear correlation in quantum dynamics. 
	
	With the DVR set, the molecular wave function is given by 
	$
	\Psi(\mathbf{r}, \mathbf{q}, t) = \sum_{\mathbf{n},\alpha} C_{\mathbf{n}\alpha}(t)\, \phi_\alpha(\mathbf{r}; \mathbf{q}_\mathbf{n})\, \chi_\mathbf{n}(\mathbf{q}),
	$
	where $\chi_\mathbf{n}(\mathbf{q})$ are the DVR basis functions for the nuclear degrees of freedom,
	and the equation of motion becomes 
	\begin{equation}\label{eq:ldr_eom}
		i\dot{C}_{\mathbf{m}\beta}(t) = V_{\mathbf{m}\beta}\, C_{\mathbf{m}\beta}(t) + \sum_{\mathbf{n},\alpha} T_\mathbf{mn}\, A^{\beta\alpha}_\mathbf{mn}\, C_{\mathbf{n}\alpha}(t),
	\end{equation}
	where $V_{\mathbf{m}\beta} = V_\beta(\mathbf{q}_\mathbf{m})$ is the electronic energy at grid point $\mathbf{q}_\mathbf{m}$ and $T_\mathbf{mn} = \langle \chi_\mathbf{m} \vert \hat{T}_{\mathrm{N}} \vert \chi_\mathbf{n} \rangle$ is the nuclear kinetic energy matrix element in the DVR basis.
	See Computational Details for further information.
	
\begin{figure*}[t]
		\centering
		\includegraphics[width=1.0\textwidth]{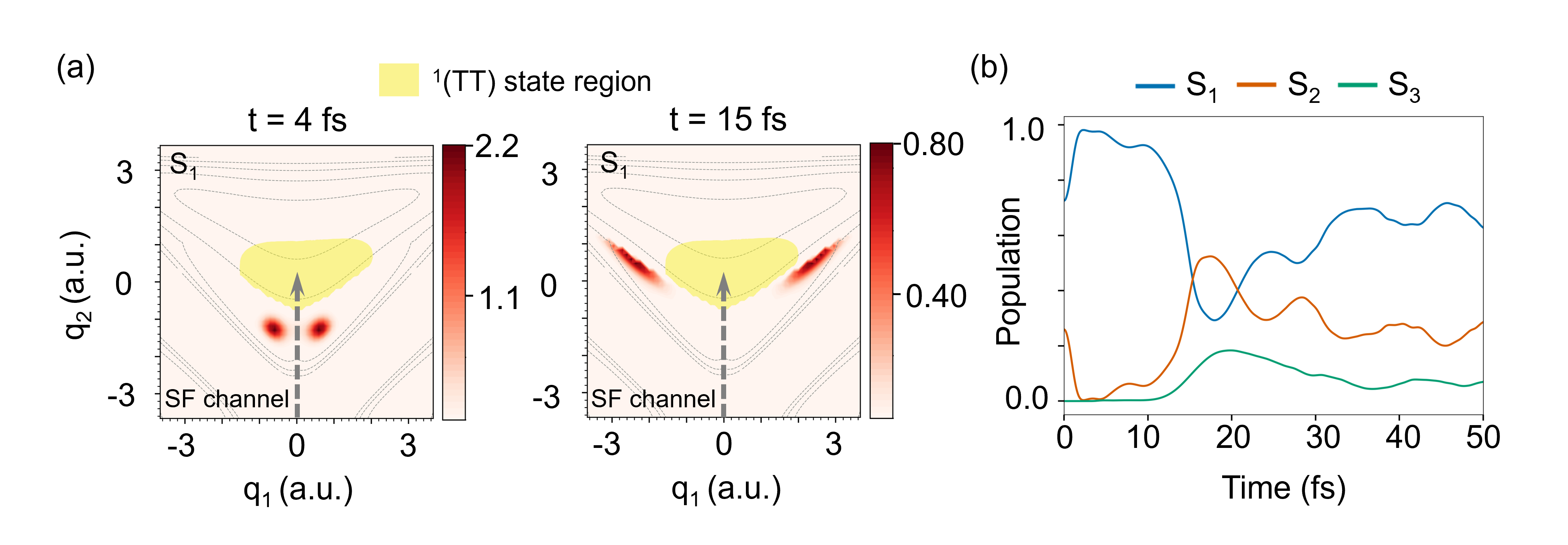}
		\caption{CSB dynamics of the H$_4$ chain ($R_{\mathrm{H1-H4}}=11.3$ Bohr).  
			(a) Distribution of nuclear wave packets on the S$_1$ state at 4 fs and 15 fs, with the light-yellow shaded area indicating the $^1(\mathrm{TT})$ region;  
			(b) Electronic population dynamics of the excited states S$_1$, S$_2$, and S$_3$ up to 50 fs.}
		\label{fig:dynamics}
	\end{figure*}

	Surprisingly, the nonadiabatic quantum dynamics simulations show that the SF reaction does \emph{not} occur in this system. Upon a vertical excitation to the bright S$_1$ state from the ground state ($R_{\mathrm{H1{-}H2}} = R_{\mathrm{H3{-}H4}} = 1.42$~Bohr and $R_{\mathrm{H2{-}H3}} = 8.50$~Bohr at equilibrium), the nuclear wavepacket initially propagates along the $q_2 > 0$ direction, encountering the S$_1$--S$_2$ crossing seam at approximately 10~fs (\cref{fig:dynamics}a).
	During 10--15~fs, the wavepacket travels along the seam, accompanied by strong nonadiabatic transitions take place: the S$_1$ population decreases rapidly within a few femtoseconds, while population transfer to higher excited states occur through the crossing seam. The dynamics results indicate that starting from two weakly interacting H$_2$ molecules, both H--H bonds are stretched but symmetry is broken, with the H$_4$ chain undergoing a one-sided H$_2$ bond dissociation process. 
	After approximately 30~fs, the electronic population distribution reaches a quasi-steady state that remains localized in the vicinity of the crossing seam, with no distribution observed in the energetically lower $^1(\mathrm{TT})$ region (\cref{fig:dynamics}b). These dynamical results presented above demonstrate that the SF channel is blocked. But how does this occur?

	\subsubsection{Is it due to the barrier?}
	
\begin{figure*}[t]
		\centering
		\includegraphics[width=1.0\textwidth]{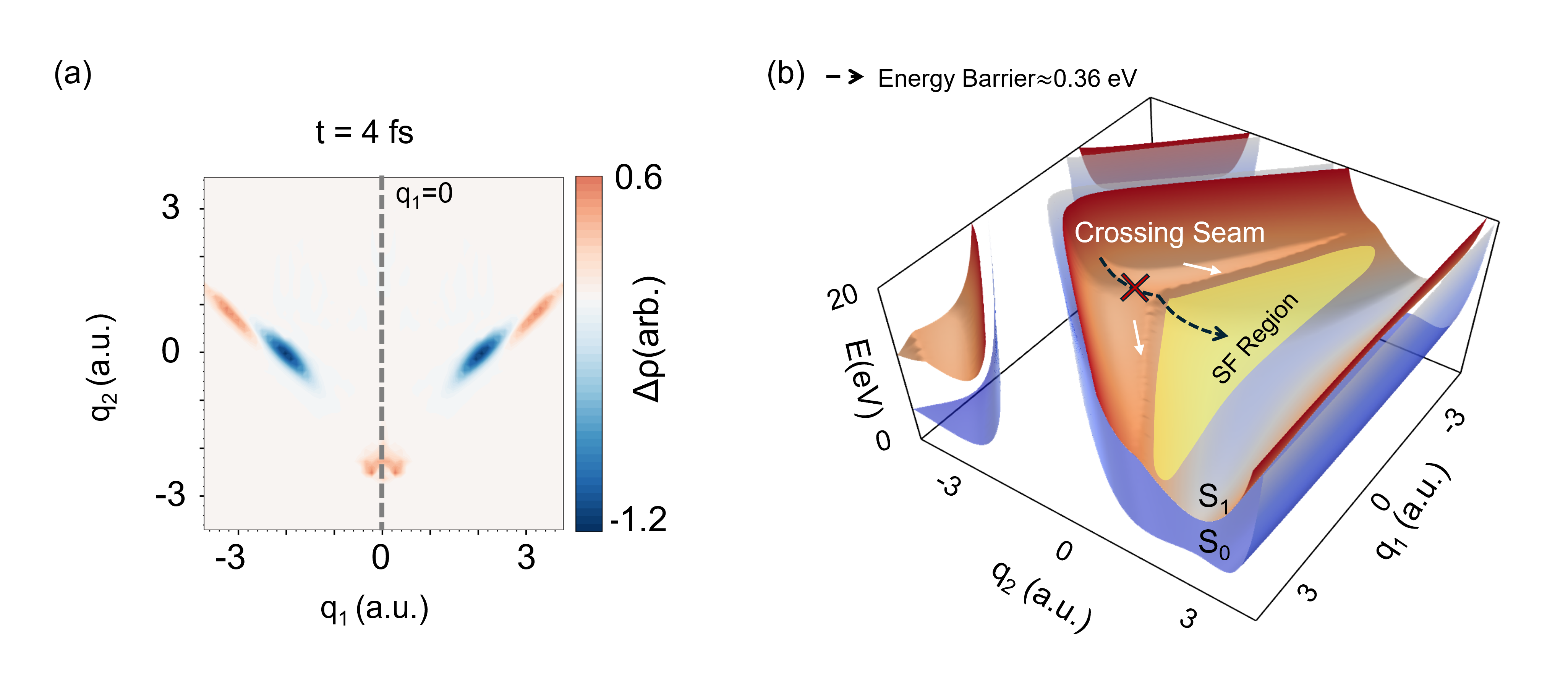}
		\caption{Nonadiabatic molecular quantum dynamics with excess nuclear kinetic energy.  
			(a) Differential nuclear density on the S$_1$ state at 10 fs, comparing simulations with initial kinetic energies of 0 eV and 2.96 eV. The density difference $\Delta\rho = \rho_{\mathrm{2.96\,eV}} - \rho_{\mathrm{0\,eV}}$ exhibits a localized redistribution pattern along the crossing seam, with no significant accumulation in the TT region ($q_2 > 0$).  
			(b) APES of the S$_0$ and S$_1$ states in the $(q_1, q_2)$ coordinate space, illustrating the energy barrier on the S$_1$ surface along the pathway toward the $^1(\mathrm{TT})$ region. The barrier height is approximately 0.36~eV, far below the injected kinetic energy of 2.96~eV.}
		\label{fig:barrier}
	\end{figure*}

	A natural question is whether this blockade simply arises from an energetic barrier on the APES. Along the $q_1 = 0$ coordinate, the barrier height required to access the $^1(\mathrm{TT})$ region is about 0.36~eV (\cref{fig:barrier}b). To test this possibility, we performed additional dynamics simulations in increasing the nuclear kinetic energy along the reaction coordinate with $\mathbf{p}_0 = (0, 20)$~a.u., corresponding to an  initial kinetic energy of 2.96~eV---far exceeding the barrier height. If the confinement were simply due to the energy barrier, the wavepacket should  overcome this barrier and enter the $^1(\mathrm{TT})$ region.
	
	Even with a large kinetic energy that far exceeds the barrier height, the wavepacket still does not traverse the crossing seam and arrive at the $^1(\mathrm{TT})$ region. Comparison of the nuclear density on the S$_1$ state at $t = 10$~fs with and without the injected kinetic energy reveals that the wavepacket simply reaches the crossing seam faster rather than overcoming the barrier to open the SF channel (\cref{fig:barrier}a). The density difference is along the crossing seam direction without any distribution observed in the $^1(\mathrm{TT})$ region. This result demonstrates that the blockade does not originate from the barrier.

	\subsubsection{Is it due to the geometric phase effect?}
	
\begin{figure}[t]
		\centering
		\includegraphics[width=0.8\textwidth]{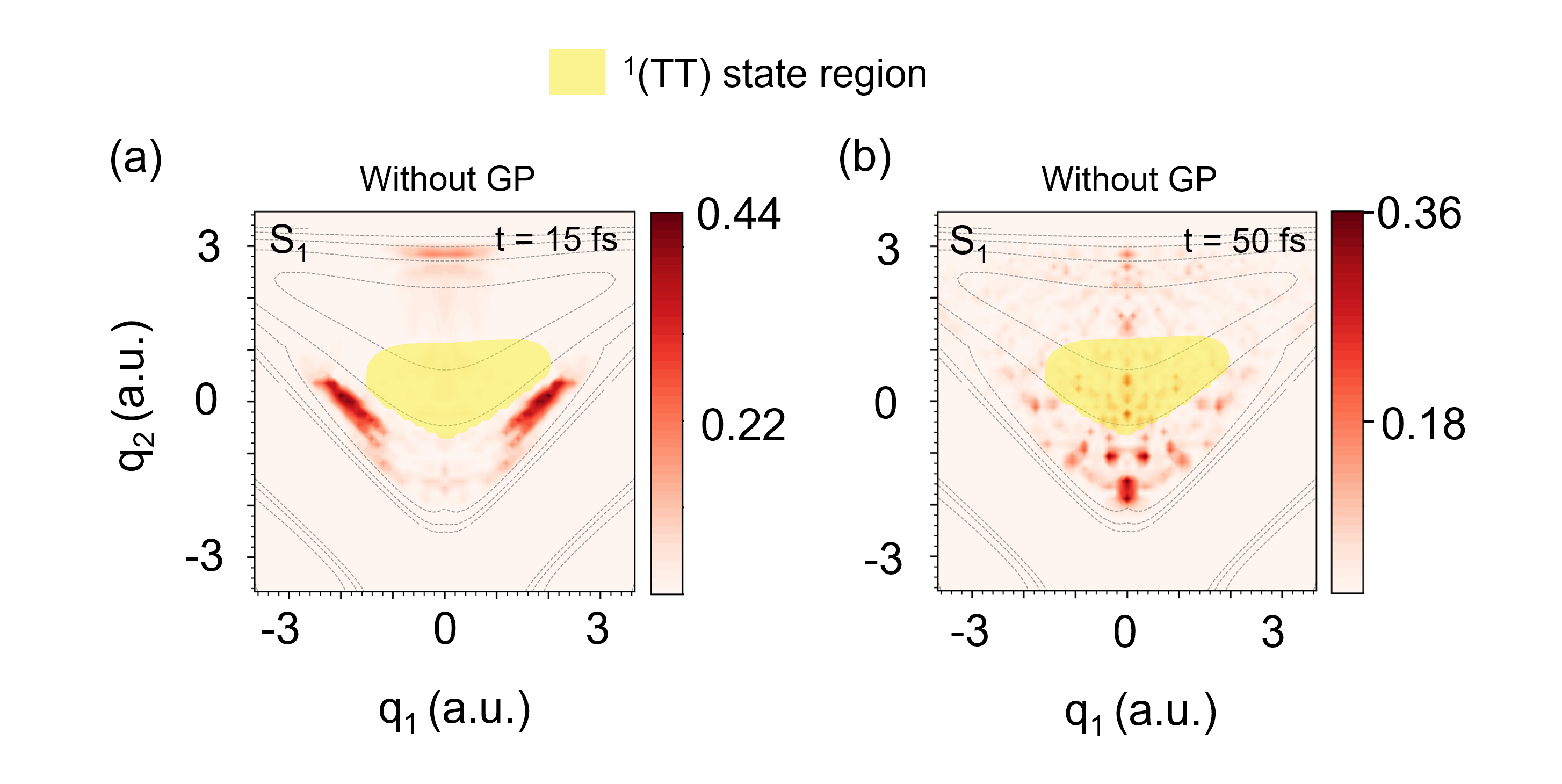}
		\caption{Influence of geometric phase in the SF dynamics. 
			(a) Nuclear density distribution on the S$_1$ state at 15~fs without geometric phase correction. 
			(b) Nuclear density distribution on the S$_1$ state at 50~fs without geometric phase correction. The light-yellow shaded area indicates the region of triplet-pair character. Removal of geometric phase allows limited density to enter the $^1(\mathrm{TT})$ region, demonstrating that geometric phase contributes to the blockade but cannot fully account for the complete closure of the SF channel.}
		\label{fig:gp}
	\end{figure}

	Another possible mechanism for the blockade is the geometric phase effects. In the presence of a conical intersection, two nuclear wavepackets can exhibit destructive quantum interference due to the geometric phase, which have been demonstrated to play significant roles in various photochemical reactions \cite{vonbuschUnambiguousProofBerrys1998,xieNonadiabaticTunnelingPhotodissociation2016b,yuanObservationGeometricPhase2018}.
	To assess the influence of geometric phase on the blockade, we performed quantum dynamics simulations with the geometric phase correction artificially removed. 
	One advantage of the geometric quantum dynamics is that it is, in fact, straightforward to turn off the geometric phase but still keeping nonadiabatic transitions intact. 
	This is because the geometric phase is encoded in the phase of the electronic overlap matrix.
	By removing the phase in all electronic overlap matrix elements, i.e.,  
	\be A^{\beta\alpha}_\mathbf{mn} \to |A^{\beta\alpha}_\mathbf{mn}|, 
	\ee
	the relative phase between electronic states is discarded, thus turning off the geometric phase effect in the nuclear dynamics. 
	As shown in \cref{fig:gp}a, the dynamics without the geometric phase show that the early-time dynamics is not affected; only at $t = 50$~fs (\cref{fig:gp}b) we observe a small portion of the density entering the $^1(\mathrm{TT})$ region.
	Therefore, the geometric phase effect, while does influence dynamics at longer time scales, cannot fully explain the blockade phenomenon. 
	
	\subsection{Quantum Geometric Picture for CSB}
	
\begin{figure}[t]
		\centering
		\includegraphics[width=0.8\textwidth]{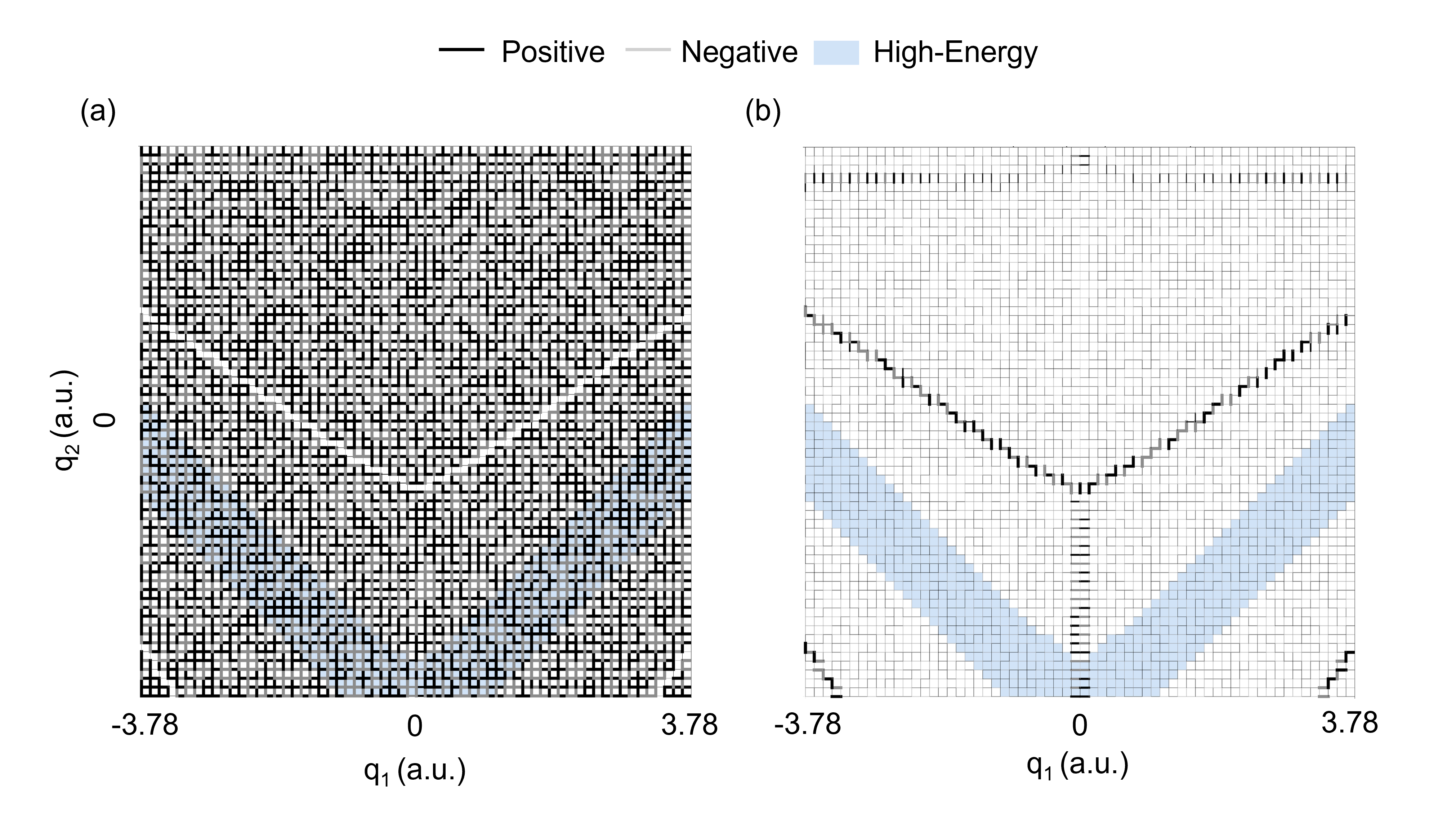}
		\caption{Geometric picture for the CSB.  
			(a) Electronic intrastate overlap matrix of the S$_1$ state on the ($q_1$, $q_2$) grid. Regions with overlap values approaching zero indicate abrupt changes in electronic state composition, coinciding with the energy near-degeneracy region on the APES.  
			(b) Electronic interstate overlap matrix between S$_1$ and S$_2$ states. Significant values at the crossing seam region confirm the abrupt character exchange between the two states, demonstrating that the crossing seam serves as both a blockade for the SF channel and a pathway for nonadiabatic transitions.}
		\label{fig:overlap}
	\end{figure}

	We now provide a quantum geometric picture to understand the blockade, in which  the primary cause of the CSB lies in the abrupt change of electronic state character across the crossing seam.
	This is reflected in the  nearest-neighbor S$_1$  intrastate electronic overlap matrix elements (\cref{fig:overlap}a), measuring the similarity between many-electron wavefunctions at different geometries\cite{xieLinkedProductApproximation2025}. 
	It is usually expected that the intrastate overlap between adjacent structures on the same adiabatic electronic state should be close to unity. 
	However, along the crossing seam, we find that the electronic overlap drops significantly to nearly zero. 
	From the electronic structure analysis,  the electronic character of geometries A and B shows a transformation from a single-excitation dominated character (geometry A) to a double-excitation-dominated character (geometry B, double-excitation contribution $\sim 70\%$). This abrupt change occurs along the entire seam, thus any  intrastate overlap matrix  elements between the reactant region and the product regions vanish. Therefore, the electronic-overlap-dressed kinetic energy matrix elements vanishes, inhibiting the singlet fission channel.

	Since the crossing seam arises from S$_1$--S$_2$ near-degeneracy, the same 
	near-zero nearest-neighbor intrastate overlap is observed for S$_2$ 
	(Fig.\ S4), consistent with our observation that the 
	crossing seam originates from electronic near-degeneracy between S$_1$ and 
	S$_2$ on the APES.
	Note that electronic near-degeneracy alone does not necessarily mean there is abrupt change of electronic character.  
	In the region with $q_1=0$ and $q_2<0$, S$_1$ and S$_2$ are 
	also nearly degenerate (see \cref{fig:pathways}a for the energy gap), yet the intrastate overlap does not vanish. 
	In this region, the intrastate overlap does not collapse to zero. This comparison indicates that the essential condition for CSB is the abrupt change of 
	electronic state character across adjacent geometries, as manifested by 
	the vanishing nearest-neighbor intrastate overlap, rather than 
	near-degeneracy itself.
	
	Furthermore, the S$_1$--S$_2$ inter-state overlap deviates far from 0 (\cref{fig:overlap}b), demonstrating that the abrupt electronic state change at the 
	crossing seam not only blocks the SF reaction channel but also serves as a 
	pathway for nonadiabatic transitions, consistent with the dynamics 
	simulation results.
	
	When the nuclear wavepacket reaches this region during the quantum dynamics, the 
	above observation directly explains the CSB mechanism. The molecule cannot 
	continue along the anticipated SF channel toward the $^1(\mathrm{TT})$ 
	region; instead, it travels along the crossing seam where the electronic states are similar and the undergoes strong nonadiabatic transitions (\cref{fig:dynamics}b).

	\subsection{Ehrenfest Dynamics Cannot Describe the CSB}
	
	It is instructive to examine whether the CSB is intrinsically a nuclear quantum effect,  requiring fully quantum dynamics that properly describe wavepacket interference and bifurcation. Here we compare  the mixed quantum-classical Ehrenfest dynamics to the exact results, which shows that Ehrenfest trajectories can in fact traverse the crossing seam region and enter the $^1(\mathrm{TT})$ region (Fig.\ S5). This demonstrates that treating nuclear motion with classical trajectories evolving on a mean-field potential cannot faithfully represent quantum interference and wavepacket bifurcation in the vicinity of electronic degeneracies, thus highlights the necessity of fully quantum dynamics methods for accurately describing the CSB phenomenon. In fact, we expect that the CSB can be challenging for other mixed quantum-classical methods such as trajectory surface-hopping \cite{tullyMixedQuantumClassical1998}.  
	
	\subsection{Modulating the Crossing Seam}
	
	As the triplet-pair state exits for a wide range of interatomic distances, we investigate how the crossing seam alters as the chain length. 
	As the terminal separation decreases, the global crossing seam on the APES progressively contracts, becoming a single conical intersection point between S$_1$ and S$_2$ at $(q_1, q_2) \approx (0,\,-0.4)$~Bohr, and eventually vanishes entirely as the two adiabatic surfaces become fully separated. This sensitive dependence of the crossing seam on chain length may offer a means to modulate the blockade strength and thereby control access to photochemical reaction channels.
	
	At the chain length of $R_{\mathrm{H1{-}H4}} = 9.0$~Bohr the APES exhibits partial contraction of the degeneracy region while retaining a segment of the crossing seam (Fig.\ S6). The S$_3$ surface separates further from the lower two adiabatic states, yet the remnant crossing seam continues to mediate nonadiabatic transitions. Fragment local spin analysis yields a global maximum of 1.82, indicating spin polarization that deviates from the singlet baseline but does not reach the value of 2 characteristic of  a triplet-pair fragments (Fig.\ S7a). 
	The quantum dynamics reveal the consequence of this partial blockade: at 6~fs, the wavepacket encounters the remaining crossing seam and is initially impeded from entering the product region; however, by 15~fs, the nuclear density circumvents the attenuated blockade and penetrates into the reactive region, with appreciable distribution persisting at 40~fs (Fig.\ S7c). The population dynamics confirm that the residual crossing seam still functions as a nonadiabatic transition pathway, facilitating S$_1 \to$ S$_2$ transfer (Fig.\ S7b).

	When the terminal separation is further reduced to $R_{\mathrm{H1{-}H4}} = 7.2$~Bohr the crossing seam shrinks to a localized CI at $(q_1, q_2) \approx (0,\,-0.4)$~Bohr (Fig.\ S8). The spatial extent of the degeneracy is markedly reduced, and the adiabatic electronic states become more distinctly separated, confining nonadiabatic effects to a compact region of configuration space. Local spin analysis of this short H$_4$ chain reveals a qualitative difference in the static electronic structure relative to the crossing seam system. Although the fragment local spin deviates from the singlet baseline in certain regions, indicating a degree of spin polarization and localization tendency, its magnitude does not reach values consistent with stable triplet fragments (Fig.\ S9a). In contrast to the crossing seam system, where a distinct $^1(\mathrm{TT})$ region can be clearly identified from the local spin signatures, the short-chain system lacks definitive static $^1(\mathrm{TT})$ indicators. Accordingly, no well-defined potential pathway corresponding to a SF reaction channel is present.
	
	The CI produces quantum dynamics distinct from the crossing seam. At 4~fs, the wavepacket propagates along the $q_1 = 0$ direction and reaches the CI region, where strong nonadiabatic transitions occur within this electronics degeneracy. By 10~fs, a portion of the nuclear density bypasses the CI and continues to propagate toward $q_2 > 0$, while the overall density on S$_1$ decreases substantially; by 30~fs, the wavepacket has largely dispersed beyond the CI region Fig.\ S9c). The population dynamics show that internal conversion from S$_1$ to S$_2$ completes on an ultrafast timescale ($<5$~fs), accompanied by CI-driven S$_1$ and S$_2$ coherent exchange (Fig.\ S9b). Since S$_3$ maintains a significant energy gap from S$_1$ and S$_2$ without electronic degeneracy, it does not constitute a transition channel, and essentially no S$_3$ distribution appears during the entire 50~fs evolution.
	
	The electronic intrastate overlap matrix provides direct confirmation of the electronic degeneracy topology in these two systems. For the intermediate chain ($R_{\mathrm{H1{-}H4}} = 9.0$~Bohr), the S$_1$ intrastate overlap matrix shows that the crossing seam has partially contracted compared to the longer chain: the region of abrupt overlap change is spatially reduced, and the magnitude of the overlap reduction within this region is attenuated, while the remainder of configuration space maintains continuous overlap values Fig.\ S11a). For the short chain ($R_{\mathrm{H1{-}H4}} = 7.2$~Bohr), according to the geometric phase theory, an adiabatic electronic wavefunction accumulates a phase of $\pi$ upon encircling a CI, which manifests as a negative loop value in the electronic intrastate overlap matrix. This verifies that the CI between S$_1$ and S$_2$ is located near $(q_1, q_2) \approx (0,\,-0.4)$~Bohr (Fig.\ S11b). Apart from this typical CI signature, the global intrastate overlap matrix values remain continuous without discontinuities, confirming the complete disappearance of the extended crossing seam.
	
	When the terminal separation is further reduced to $R_{\mathrm{H1{-}H4}} = 5.0$~Bohr (Fig.\ S10), the S$_1$ and S$_2$ APESs become completely separated, with neither near-degeneracy nor conical intersection present in the explored configuration space. In this regime, the electronic states evolve smoothly and remain energetically distinct, eliminating the possibility of strong nonadiabatic coupling. The absence of electronic degeneracy implies that transitions between S$_1$ and S$_2$ are blocked due to the energy gap. The corresponding electronic intrastate overlap matrix (Fig.\ S11c) shows no reduction overlap regions. The intrastate overlap between adjacent geometries remains close to unity throughout the configuration space, consistent with smoothly varying adiabatic electronic states on well-separated potential energy surfaces. 
	
	These results demonstrate that the chain length systematically modulates the crossing seam topology: from a global crossing seam that enforces complete dynamical blockade, through partial seam contraction that weakens the blockade and permits reaction bypass, to a localized CI that enables ultrafast nonadiabatic exchange without CSB, and ultimately to a fully separated adiabatic regime in which electronic degeneracy disappears.
	Importantly, the emergence of a well-defined $^1(\mathrm{TT})$ region is intimately linked to the extended crossing seam topology. When the seam contracts to a localized CI, or vanishes entirely as the adiabatic surfaces separate, the triplet-pair character no longer contributes to S$_1$. 
	
	In the presence of a crossing seam, it should be emphasized that the nuclear motion is entirely dictated by the electronic quantum geometry, instead of the topology of the potential energy surfaces. 
	This tunability suggests that molecular design strategies targeting the crossing seam extent could provide a mechanism for controlling photochemical reaction pathways.

	\section{Computational Details}\label{sec:comp_details}
	
	\textbf{Electronic Structure Calculations.}
	All electronic structure calculations were performed using PySCF 
	\cite{sunRecentDevelopmentsSCF2020}. 
	The atomic motion is described by two collective reaction coordinates: the 
	symmetric stretch mode $q_1$ and the antisymmetric stretch mode $q_2$, 
	defined through
	$
	\Delta x_2 = \frac{1}{\sqrt{2}}(q_1 + q_2), 
	\Delta x_3 = \frac{1}{\sqrt{2}}(q_1 - q_2),
	$
	with $x_2 = x_2^{0} + \Delta x_2$ and $x_3 = x_3^{0} + \Delta x_3$.
	In the $(q_1, q_2)$ coordinate space, a two-dimensional potential energy 
	surface  was constructed spanning the range $[-3.78,\, 3.78]$~Bohr 
	with a uniform $64 \times 64$ grid. At each grid point, the energies and 
	wavefunctions of the lowest four singlet states (S$_0$--S$_3$) were computed 
	at the FCI/cc-pVDZ level of theory.
	
	\textbf{Exact Quantum Dynamics.}
	All quantum dynamics codes are implemented in the open-source Python-based program \textsc{PyQED} \cite{xiePyQEDPythonFramework2026}. The sine DVR basis sets were employed for the nuclear degrees of freedom over the same grid range and resolution as the APES construction described above. The initial nuclear wavepacket was prepared as a two-dimensional Gaussian wave packet
	\begin{equation}\label{eq:gwp}
		\chi_0(q_1, q_2) = \frac{1}{\sqrt{\pi\,\sigma_{1}\sigma_{2}}}
		\exp\!\left(-\frac{(q_1 - q_{1}^0)^2}{2\sigma_{1}^2} 
		- \frac{(q_2 - q_{2}^0)^2}{2\sigma_{2}^2}\right),
	\end{equation}
	centered at $(q_{1}^0,\, q_{2}^0) = (0,\, -2)$~Bohr on the S$_1$ 
	state, corresponding to the Franck--Condon point upon vertical excitation 
	from the ground-state equilibrium geometry. The  width $\sigma_1 = 
	\sigma_2 = 0.193$~Bohr are determined from a harmonic analysis of the 
	S$_0$ potential energy surface at the equilibrium geometry. 
	The equations of motion (Eq.~\eqref{eq:ldr_eom}) were integrated with a time step of $\Delta t = 0.001$~fs.
	The computational details with different terminal separations 
	$R_{\mathrm{H1-H4}}$ are provided in the Supporting Information.
	
	\textbf{Ehrenfest Dynamics.}
	Mixed quantum-classical Ehrenfest dynamics simulations were performed using 500 independent trajectories, all initiated on the S$_1$ state. The initial nuclear positions were sampled from the Gaussian wave packet in Eq.~\eqref{eq:gwp}, and the initial momenta were set to zero for all trajectories. Electronic energies, gradients, and nonadiabatic couplings were computed on-the-fly at the SA-CASSCF(4,4)/cc-pVDZ level, using equal weighting over the lowest four singlet states. The nuclear motion was propagated with the velocity Verlet algorithm using a time step of $\Delta t = 0.1$~fs for a total simulation time of 50~fs.
	
	\section{Conclusions}
	
	This study reveals a photochemical phenomenon governed by the electronic near-degeneracy topology of the excited-state manifold, which we term the crossing seam blockade. Using a H$_4$ chain molecule, we show that although the system satisfies the SF energy criterion and exhibits well-defined static triplet-pair local spin signatures consistent with a $^1(\mathrm{TT})$ character in specific regions of nuclear configuration space, the actual post-excitation evolution does not proceed along the anticipated SF channel. Full quantum dynamics simulations demonstrate that the nuclear wavepacket is trapped in the vicinity of the S$_1$--S$_2$ crossing seam, where strong nonadiabatic transitions and population redistribution occur. As a result, the system fails to penetrate the degeneracy seam and access the $^1(\mathrm{TT})$ region. Notably, this blockade persists even when an initial kinetic energy far exceeding the nominal barrier height is injected, ruling out conventional energy-controlled barrier crossing as the underlying mechanism. We further examined the role of geometric phase effects, which contribute to reinforcing the blockade at longer evolution times by enhancing quantum interference, though they alone cannot fully account for the complete closure of the SF channel. The geometric picture for the CSB is ultimately provided by the electronic intrastate overlap matrix analysis, which reveals an abrupt change of electronic state character across the crossing seam that completely blocks the reaction pathway.
	
	These findings have important implications for understanding SF and, more broadly, photochemical reaction mechanisms. Electronic degeneracy structures are often assumed to facilitate photochemical reactivity by providing efficient nonradiative funnels. In contrast, our results demonstrate that electronic degeneracy and near-degeneracy do not necessarily open the expected reaction channels; instead, they can suppress specific pathways by inducing strong electronic state mixing and nonadiabatic population transfer in the degeneracy region. Importantly, we show that the crossing seam topology can be systematically modulated by varying the H chain length: as the terminal separation decreases, the extended seam progressively contracts, first weakening the blockade to permit partial reaction bypass at intermediate chain lengths, and ultimately collapsing into a localized CI where ultrafast nonadiabatic transitions occur without a blockade effect. However, in the shortest chain, the absence of stable static $^1(\mathrm{TT})$ signatures still prevents the formation of an effective SF channel. This tunability suggests that the CSB represents not only a distinct photochemical mechanism but also a potential control handle for modulating energy-conversion pathways through molecular design strategies that target the crossing seam extent.
	
	From a methodological standpoint, we have demonstrated that the quantum geometrical molecular dynamics framework, regularized by the discrete variable representation basis set, can not only describe conical intersection dynamics, but also nonadiabatic dynamics in the presence of a crossing seam, accurately capturing quantum interference and wavepacket bifurcation in these regions whereby mixed quantum-classical Ehrenfest dynamics fails to capture the CSB. This suggests the necessity of full quantum dynamics methods when addressing photochemical processes with complex electronic quantum geometry.

	\section*{Supporting Information}
	This PDF file includes Fig.\ S1 to S11: APES at various terminal separations (Figs.\ S1, S6, S8, S10), comparison of ground-state surfaces at different levels of theory (Fig.\ S2), energy criterion maps for SF (Fig.\ S3), electronic intrastate overlap matrices (Figs.\ S4, S11), Ehrenfest dynamics trajectories and population evolution (Fig.\ S5), and quantum dynamics with local spin analysis at different terminal separations (Figs.\ S7, S9).
	\bibliography{H4}

\providecommand{\latin}[1]{#1}
\makeatletter
\providecommand{\doi}
  {\begingroup\let\do\@makeother\dospecials
  \catcode`\{=1 \catcode`\}=2 \doi@aux}
\providecommand{\doi@aux}[1]{\endgroup\texttt{#1}}
\makeatother
\providecommand*\mcitethebibliography{\thebibliography}
\csname @ifundefined\endcsname{endmcitethebibliography}
  {\let\endmcitethebibliography\endthebibliography}{}
\begin{mcitethebibliography}{67}
\providecommand*\natexlab[1]{#1}
\providecommand*\mciteSetBstSublistMode[1]{}
\providecommand*\mciteSetBstMaxWidthForm[2]{}
\providecommand*\mciteBstWouldAddEndPuncttrue
  {\def\EndOfBibitem{\unskip.}}
\providecommand*\mciteBstWouldAddEndPunctfalse
  {\let\EndOfBibitem\relax}
\providecommand*\mciteSetBstMidEndSepPunct[3]{}
\providecommand*\mciteSetBstSublistLabelBeginEnd[3]{}
\providecommand*\EndOfBibitem{}
\mciteSetBstSublistMode{f}
\mciteSetBstMaxWidthForm{subitem}{(\alph{mcitesubitemcount})}
\mciteSetBstSublistLabelBeginEnd
  {\mcitemaxwidthsubitemform\space}
  {\relax}
  {\relax}

\bibitem[Domcke \latin{et~al.}(2011)Domcke, Yarkony, and
  K{\"o}ppel]{domckeConicalIntersectionsTheory2011}
Domcke,~W.; Yarkony,~D.~R.; K{\"o}ppel,~H. \emph{Conical {{Intersections}}:
  {{Theory}}, {{Computation}} and {{Experiment}}}; Advanced {{Series}} in
  {{Physical Chemistry}}; WORLD SCIENTIFIC, 2011; Vol.~17\relax
\mciteBstWouldAddEndPuncttrue
\mciteSetBstMidEndSepPunct{\mcitedefaultmidpunct}
{\mcitedefaultendpunct}{\mcitedefaultseppunct}\relax
\EndOfBibitem
\bibitem[Baer and Billing(2002)Baer, and
  Billing]{baerAdvancesChemicalPhysics2002}
Baer,~M., Billing,~G.~D., Eds. \emph{Advances in Chemical Physics: {{Volume}}
  124: {{The}} Role of Degenerate States in Chemistry}; Advances in Chemical
  Physics v. 124; J. Wiley \& Sons: Hoboken, N.J, 2002\relax
\mciteBstWouldAddEndPuncttrue
\mciteSetBstMidEndSepPunct{\mcitedefaultmidpunct}
{\mcitedefaultendpunct}{\mcitedefaultseppunct}\relax
\EndOfBibitem
\bibitem[Nakamura(2012)]{nakamuraNonadiabaticTransitionConcepts2012}
Nakamura,~H. \emph{Nonadiabatic {{Transition}}: {{Concepts}}, {{Basic
  Theories}} and {{Applications}}}, 2nd ed.; WORLD SCIENTIFIC, 2012\relax
\mciteBstWouldAddEndPuncttrue
\mciteSetBstMidEndSepPunct{\mcitedefaultmidpunct}
{\mcitedefaultendpunct}{\mcitedefaultseppunct}\relax
\EndOfBibitem
\bibitem[Domcke and Yarkony(2012)Domcke, and
  Yarkony]{domckeRoleConicalIntersections2012}
Domcke,~W.; Yarkony,~D.~R. Role of {{Conical Intersections}} in {{Molecular
  Spectroscopy}} and {{Photoinduced Chemical Dynamics}}. \emph{Annual Review of
  Physical Chemistry} \textbf{2012}, \emph{63}, 325--352\relax
\mciteBstWouldAddEndPuncttrue
\mciteSetBstMidEndSepPunct{\mcitedefaultmidpunct}
{\mcitedefaultendpunct}{\mcitedefaultseppunct}\relax
\EndOfBibitem
\bibitem[Born and Oppenheimer(1927)Born, and
  Oppenheimer]{bornZurQuantentheorieMolekeln1927}
Born,~M.; Oppenheimer,~R. Zur {{Quantentheorie}} Der {{Molekeln}}.
  \emph{Annalen der Physik} \textbf{1927}, \emph{389}, 457--484\relax
\mciteBstWouldAddEndPuncttrue
\mciteSetBstMidEndSepPunct{\mcitedefaultmidpunct}
{\mcitedefaultendpunct}{\mcitedefaultseppunct}\relax
\EndOfBibitem
\bibitem[Tannor(2007)]{tannorIntroductionQuantumMechanics2007}
Tannor,~D.~J. \emph{Introduction to Quantum Mechanics: A Time-Dependent
  Perspective}; University Science Books: Sausalito, Calif, 2007\relax
\mciteBstWouldAddEndPuncttrue
\mciteSetBstMidEndSepPunct{\mcitedefaultmidpunct}
{\mcitedefaultendpunct}{\mcitedefaultseppunct}\relax
\EndOfBibitem
\bibitem[Kandori \latin{et~al.}(2001)Kandori, Shichida, and
  Yoshizawa]{kandoriPhotoisomerizationRhodopsin2001}
Kandori,~H.; Shichida,~Y.; Yoshizawa,~T. Photoisomerization in {{Rhodopsin}}.
  \emph{Biochemistry (Moscow)} \textbf{2001}, \emph{66}, 1197--1209\relax
\mciteBstWouldAddEndPuncttrue
\mciteSetBstMidEndSepPunct{\mcitedefaultmidpunct}
{\mcitedefaultendpunct}{\mcitedefaultseppunct}\relax
\EndOfBibitem
\bibitem[Polli \latin{et~al.}(2010)Polli, Alto{\`e}, Weingart, Spillane,
  Manzoni, Brida, Tomasello, Orlandi, Kukura, Mathies, Garavelli, and
  Cerullo]{polliConicalIntersectionDynamics2010}
Polli,~D.; Alto{\`e},~P.; Weingart,~O.; Spillane,~K.~M.; Manzoni,~C.;
  Brida,~D.; Tomasello,~G.; Orlandi,~G.; Kukura,~P.; Mathies,~R.~A.;
  Garavelli,~M.; Cerullo,~G. Conical Intersection Dynamics of the Primary
  Photoisomerization Event in Vision. \emph{Nature} \textbf{2010}, \emph{467},
  440--443\relax
\mciteBstWouldAddEndPuncttrue
\mciteSetBstMidEndSepPunct{\mcitedefaultmidpunct}
{\mcitedefaultendpunct}{\mcitedefaultseppunct}\relax
\EndOfBibitem
\bibitem[Barbatti \latin{et~al.}(2010)Barbatti, Aquino, Szymczak,
  Nachtigallov{\'a}, Hobza, and
  Lischka]{barbattiRelaxationMechanismsUVphotoexcited2010}
Barbatti,~M.; Aquino,~A. J.~A.; Szymczak,~J.~J.; Nachtigallov{\'a},~D.;
  Hobza,~P.; Lischka,~H. Relaxation Mechanisms of {{UV-photoexcited DNA}} and
  {{RNA}} Nucleobases. \emph{Proceedings of the National Academy of Sciences}
  \textbf{2010}, \emph{107}, 21453--21458\relax
\mciteBstWouldAddEndPuncttrue
\mciteSetBstMidEndSepPunct{\mcitedefaultmidpunct}
{\mcitedefaultendpunct}{\mcitedefaultseppunct}\relax
\EndOfBibitem
\bibitem[Prokhorenko \latin{et~al.}(2016)Prokhorenko, Picchiotti, Pola,
  Dijkstra, and Miller]{prokhorenkoNewInsightsPhotophysics2016}
Prokhorenko,~V.~I.; Picchiotti,~A.; Pola,~M.; Dijkstra,~A.~G.; Miller,~R. J.~D.
  New {{Insights}} into the {{Photophysics}} of {{DNA Nucleobases}}. \emph{The
  Journal of Physical Chemistry Letters} \textbf{2016}, \emph{7},
  4445--4450\relax
\mciteBstWouldAddEndPuncttrue
\mciteSetBstMidEndSepPunct{\mcitedefaultmidpunct}
{\mcitedefaultendpunct}{\mcitedefaultseppunct}\relax
\EndOfBibitem
\bibitem[Grebenshchikov \latin{et~al.}(2007)Grebenshchikov, Qu, Zhu, and
  Schinke]{grebenshchikovNewTheoreticalInvestigations2007}
Grebenshchikov,~S.~{\relax Yu}.; Qu,~Z.-W.; Zhu,~H.; Schinke,~R. New
  Theoretical Investigations of the Photodissociation of Ozone in the
  {{Hartley}}, {{Huggins}}, {{Chappuis}}, and {{Wulf}} Bands. \emph{Physical
  Chemistry Chemical Physics} \textbf{2007}, \emph{9}, 2044\relax
\mciteBstWouldAddEndPuncttrue
\mciteSetBstMidEndSepPunct{\mcitedefaultmidpunct}
{\mcitedefaultendpunct}{\mcitedefaultseppunct}\relax
\EndOfBibitem
\bibitem[Yarkony(2001)]{yarkonyIntersectingConicalIntersection2001}
Yarkony,~D.~R. Intersecting Conical Intersection Seams in Tetra-Atomic
  Molecules: The {{S}}{\textsubscript{1}} --{{S}}{\textsubscript{0}} Internal
  Conversion in {{HNCO}}. \emph{Molecular Physics} \textbf{2001}, \emph{99},
  1463--1467\relax
\mciteBstWouldAddEndPuncttrue
\mciteSetBstMidEndSepPunct{\mcitedefaultmidpunct}
{\mcitedefaultendpunct}{\mcitedefaultseppunct}\relax
\EndOfBibitem
\bibitem[Fazzi \latin{et~al.}(2012)Fazzi, Grancini, Maiuri, Brida, Cerullo, and
  Lanzani]{fazziUltrafastInternalConversion2012}
Fazzi,~D.; Grancini,~G.; Maiuri,~M.; Brida,~D.; Cerullo,~G.; Lanzani,~G.
  Ultrafast Internal Conversion in a Low Band Gap Polymer for Photovoltaics:
  Experimental and Theoretical Study. \emph{Physical Chemistry Chemical
  Physics} \textbf{2012}, \emph{14}, 6367\relax
\mciteBstWouldAddEndPuncttrue
\mciteSetBstMidEndSepPunct{\mcitedefaultmidpunct}
{\mcitedefaultendpunct}{\mcitedefaultseppunct}\relax
\EndOfBibitem
\bibitem[Xie \latin{et~al.}(2016)Xie, Ma, Zhu, Yarkony, Xie, and
  Guo]{xieNonadiabaticTunnelingPhotodissociation2016}
Xie,~C.; Ma,~J.; Zhu,~X.; Yarkony,~D.~R.; Xie,~D.; Guo,~H. Nonadiabatic
  {{Tunneling}} in {{Photodissociation}} of {{Phenol}}. \emph{Journal of the
  American Chemical Society} \textbf{2016}, \emph{138}, 7828--7831\relax
\mciteBstWouldAddEndPuncttrue
\mciteSetBstMidEndSepPunct{\mcitedefaultmidpunct}
{\mcitedefaultendpunct}{\mcitedefaultseppunct}\relax
\EndOfBibitem
\bibitem[Yang \latin{et~al.}(2016)Yang, Shen, Zhang, and
  Yang]{yangConicalIntersectionsParticle2016}
Yang,~Y.; Shen,~L.; Zhang,~D.; Yang,~W. Conical {{Intersections}} from
  {{Particle}}--{{Particle Random Phase}} and {{Tamm}}--{{Dancoff
  Approximations}}. \emph{The Journal of Physical Chemistry Letters}
  \textbf{2016}, \emph{7}, 2407--2411\relax
\mciteBstWouldAddEndPuncttrue
\mciteSetBstMidEndSepPunct{\mcitedefaultmidpunct}
{\mcitedefaultendpunct}{\mcitedefaultseppunct}\relax
\EndOfBibitem
\bibitem[Yue \latin{et~al.}(2018)Yue, Liu, and
  Zhu]{yuePerformanceTDDFTSpinflip2018}
Yue,~L.; Liu,~Y.; Zhu,~C. Performance of {{TDDFT}} with and without Spin-Flip
  in Trajectory Surface Hopping Dynamics: {\emph{Cis}} -- {\emph{Trans}}
  Azobenzene Photoisomerization. \emph{Physical Chemistry Chemical Physics}
  \textbf{2018}, \emph{20}, 24123--24139\relax
\mciteBstWouldAddEndPuncttrue
\mciteSetBstMidEndSepPunct{\mcitedefaultmidpunct}
{\mcitedefaultendpunct}{\mcitedefaultseppunct}\relax
\EndOfBibitem
\bibitem[Guan \latin{et~al.}(2021)Guan, Xie, Guo, and
  Yarkony]{guanEnablingUnifiedDescription2021}
Guan,~Y.; Xie,~C.; Guo,~H.; Yarkony,~D.~R. Enabling a {{Unified Description}}
  of {{Both Internal Conversion}} and {{Intersystem Crossing}} in
  {{Formaldehyde}}: {{A Global Coupled Quasi-Diabatic Hamiltonian}} for {{Its
  S}}{\textsubscript{0}} , {{S}}{\textsubscript{1}} , and
  {{T}}{\textsubscript{1}} {{States}}. \emph{Journal of Chemical Theory and
  Computation} \textbf{2021}, \emph{17}, 4157--4168\relax
\mciteBstWouldAddEndPuncttrue
\mciteSetBstMidEndSepPunct{\mcitedefaultmidpunct}
{\mcitedefaultendpunct}{\mcitedefaultseppunct}\relax
\EndOfBibitem
\bibitem[Hu and Huo(2023)Hu, and Huo]{huInitioMolecularCavity2023}
Hu,~D.; Huo,~P. Ab {{Initio Molecular Cavity Quantum Electrodynamics
  Simulations Using Machine Learning Models}}. \emph{Journal of Chemical Theory
  and Computation} \textbf{2023}, \emph{19}, 2353--2368\relax
\mciteBstWouldAddEndPuncttrue
\mciteSetBstMidEndSepPunct{\mcitedefaultmidpunct}
{\mcitedefaultendpunct}{\mcitedefaultseppunct}\relax
\EndOfBibitem
\bibitem[Pollak and Cao(2025)Pollak, and Cao]{pollakEffectOpticalCavity2025}
Pollak,~E.; Cao,~J. The Effect of an Optical Cavity on Diabatic Tunneling in an
  Ensemble of Symmetric Double-Well Systems. \emph{The Journal of Chemical
  Physics} \textbf{2025}, \emph{163}, 234111\relax
\mciteBstWouldAddEndPuncttrue
\mciteSetBstMidEndSepPunct{\mcitedefaultmidpunct}
{\mcitedefaultendpunct}{\mcitedefaultseppunct}\relax
\EndOfBibitem
\bibitem[Jiang \latin{et~al.}(2025)Jiang, Zhang, Wang, Peng, Jin, Zou, Zhu,
  Jiang, Lan, Yong, He, and Xiang]{jiangSuperresolutionFemtosecondElectron2025}
Jiang,~H.; Zhang,~J.; Wang,~T.; Peng,~J.; Jin,~C.; Zou,~X.; Zhu,~P.; Jiang,~T.;
  Lan,~Z.; Yong,~H.; He,~F.; Xiang,~D. Super-Resolution Femtosecond Electron
  Diffraction Reveals Electronic and Nuclear Dynamics at Conical Intersections.
  \emph{Nature Communications} \textbf{2025}, \emph{16}, 6703\relax
\mciteBstWouldAddEndPuncttrue
\mciteSetBstMidEndSepPunct{\mcitedefaultmidpunct}
{\mcitedefaultendpunct}{\mcitedefaultseppunct}\relax
\EndOfBibitem
\bibitem[Zhang \latin{et~al.}(2025)Zhang, Liu, Lin, Xu, Gu, Gelin, and
  Lan]{zhangUnderstandingMolecularMotions2025}
Zhang,~J.; Liu,~H.; Lin,~C.; Xu,~C.; Gu,~F.; Gelin,~M.~F.; Lan,~Z.
  Understanding of Molecular Motions in Nonadiabatic Photoisomerization
  Dynamics of Cis-Stilbene with on-the-Fly Simulation of Transient Absorption
  Pump--Probe Spectra. \emph{The Journal of Chemical Physics} \textbf{2025},
  \emph{163}, 244109\relax
\mciteBstWouldAddEndPuncttrue
\mciteSetBstMidEndSepPunct{\mcitedefaultmidpunct}
{\mcitedefaultendpunct}{\mcitedefaultseppunct}\relax
\EndOfBibitem
\bibitem[Duston \latin{et~al.}(2025)Duston, Bradbury, Tao, and
  Subotnik]{dustonConicalIntersectionsElectronic2025a}
Duston,~T.; Bradbury,~N.~C.; Tao,~Z.; Subotnik,~J.~E. Conical {{Intersections}}
  and {{Electronic Momentum}} as {{Viewed}} from {{Phase Space Electronic
  Structure Theory}}. \emph{The Journal of Physical Chemistry Letters}
  \textbf{2025}, \emph{16}, 8994--9003\relax
\mciteBstWouldAddEndPuncttrue
\mciteSetBstMidEndSepPunct{\mcitedefaultmidpunct}
{\mcitedefaultendpunct}{\mcitedefaultseppunct}\relax
\EndOfBibitem
\bibitem[Berry(1984)]{berryQuantalPhaseFactors1984}
Berry,~M.~V. Quantal Phase Factors Accompanying Adiabatic Changes.
  \emph{Proceedings of the Royal Society of London. A. Mathematical and
  Physical Sciences} \textbf{1984}, \emph{392}, 45--57\relax
\mciteBstWouldAddEndPuncttrue
\mciteSetBstMidEndSepPunct{\mcitedefaultmidpunct}
{\mcitedefaultendpunct}{\mcitedefaultseppunct}\relax
\EndOfBibitem
\bibitem[Mead(1992)]{meadGeometricPhaseMolecular1992}
Mead,~C.~A. The Geometric Phase in Molecular Systems. \emph{Reviews of Modern
  Physics} \textbf{1992}, \emph{64}, 51--85\relax
\mciteBstWouldAddEndPuncttrue
\mciteSetBstMidEndSepPunct{\mcitedefaultmidpunct}
{\mcitedefaultendpunct}{\mcitedefaultseppunct}\relax
\EndOfBibitem
\bibitem[Ryabinkin and Izmaylov(2013)Ryabinkin, and
  Izmaylov]{ryabinkinGeometricPhaseEffects2013a}
Ryabinkin,~I.~G.; Izmaylov,~A.~F. Geometric {{Phase Effects}} in {{Dynamics
  Near Conical Intersections}}: {{Symmetry Breaking}} and {{Spatial
  Localization}}. \emph{Physical Review Letters} \textbf{2013}, \emph{111},
  220406\relax
\mciteBstWouldAddEndPuncttrue
\mciteSetBstMidEndSepPunct{\mcitedefaultmidpunct}
{\mcitedefaultendpunct}{\mcitedefaultseppunct}\relax
\EndOfBibitem
\bibitem[Ryabinkin \latin{et~al.}(2017)Ryabinkin, {Joubert-Doriol}, and
  Izmaylov]{ryabinkinGeometricPhaseEffects2017}
Ryabinkin,~I.~G.; {Joubert-Doriol},~L.; Izmaylov,~A.~F. Geometric {{Phase
  Effects}} in {{Nonadiabatic Dynamics}} near {{Conical Intersections}}.
  \emph{Accounts of Chemical Research} \textbf{2017}, \emph{50},
  1785--1793\relax
\mciteBstWouldAddEndPuncttrue
\mciteSetBstMidEndSepPunct{\mcitedefaultmidpunct}
{\mcitedefaultendpunct}{\mcitedefaultseppunct}\relax
\EndOfBibitem
\bibitem[Farag \latin{et~al.}(2021)Farag, Mandal, and
  Huo]{faragPolaritonInducedConical2021}
Farag,~M.~H.; Mandal,~A.; Huo,~P. Polariton Induced Conical Intersection and
  Berry Phase. \emph{Physical Chemistry Chemical Physics} \textbf{2021},
  \emph{23}, 16868--16879\relax
\mciteBstWouldAddEndPuncttrue
\mciteSetBstMidEndSepPunct{\mcitedefaultmidpunct}
{\mcitedefaultendpunct}{\mcitedefaultseppunct}\relax
\EndOfBibitem
\bibitem[Handy \latin{et~al.}(1986)Handy, Yamaguchi, and
  Schaefer]{handyDiagonalCorrectionBorn1986}
Handy,~N.~C.; Yamaguchi,~Y.; Schaefer,~H.~F. The Diagonal Correction to the
  {{Born}}--{{Oppenheimer}} Approximation: {{Its}} Effect on the
  Singlet--Triplet Splitting of {{CH2}} and Other Molecular Effects. \emph{The
  Journal of Chemical Physics} \textbf{1986}, \emph{84}, 4481--4484\relax
\mciteBstWouldAddEndPuncttrue
\mciteSetBstMidEndSepPunct{\mcitedefaultmidpunct}
{\mcitedefaultendpunct}{\mcitedefaultseppunct}\relax
\EndOfBibitem
\bibitem[Baer(2006)]{baerBornOppenheimerConical2006}
Baer,~M. \emph{Beyond {{Born}}--{{Oppenheimer}}: {{Conical Intersections}} and
  {{Electronic Nonadiabatic Coupling Terms}}}, 1st ed.; Wiley, 2006\relax
\mciteBstWouldAddEndPuncttrue
\mciteSetBstMidEndSepPunct{\mcitedefaultmidpunct}
{\mcitedefaultendpunct}{\mcitedefaultseppunct}\relax
\EndOfBibitem
\bibitem[Born \latin{et~al.}(1955)Born, Huang, and
  Lax]{bornDynamicalTheoryCrystal1955}
Born,~M.; Huang,~K.; Lax,~M. Dynamical {{Theory}} of {{Crystal Lattices}}.
  \emph{American Journal of Physics} \textbf{1955}, \emph{23}, 474--474\relax
\mciteBstWouldAddEndPuncttrue
\mciteSetBstMidEndSepPunct{\mcitedefaultmidpunct}
{\mcitedefaultendpunct}{\mcitedefaultseppunct}\relax
\EndOfBibitem
\bibitem[Matselyukh \latin{et~al.}(2022)Matselyukh, Despr{\'e}, Golubev,
  Kuleff, and W{\"o}rner]{matselyukhDecoherenceRevivalAttosecond2022}
Matselyukh,~D.~T.; Despr{\'e},~V.; Golubev,~N.~V.; Kuleff,~A.~I.;
  W{\"o}rner,~H.~J. Decoherence and Revival in Attosecond Charge Migration
  Driven by Non-Adiabatic Dynamics. \emph{Nature Physics} \textbf{2022},
  \emph{18}, 1206--1213\relax
\mciteBstWouldAddEndPuncttrue
\mciteSetBstMidEndSepPunct{\mcitedefaultmidpunct}
{\mcitedefaultendpunct}{\mcitedefaultseppunct}\relax
\EndOfBibitem
\bibitem[Nakamura and Truhlar(2001)Nakamura, and
  Truhlar]{nakamuraDirectCalculationDiabatic2001}
Nakamura,~H.; Truhlar,~D.~G. The Direct Calculation of Diabatic States Based on
  Configurational Uniformity. \emph{The Journal of Chemical Physics}
  \textbf{2001}, \emph{115}, 10353--10372\relax
\mciteBstWouldAddEndPuncttrue
\mciteSetBstMidEndSepPunct{\mcitedefaultmidpunct}
{\mcitedefaultendpunct}{\mcitedefaultseppunct}\relax
\EndOfBibitem
\bibitem[Subotnik \latin{et~al.}(2008)Subotnik, Yeganeh, Cave, and
  Ratner]{subotnikConstructingDiabaticStates2008}
Subotnik,~J.~E.; Yeganeh,~S.; Cave,~R.~J.; Ratner,~M.~A. Constructing Diabatic
  States from Adiabatic States: {{Extending}} Generalized
  {{Mulliken}}--{{Hush}} to Multiple Charge Centers with {{Boys}} Localization.
  \emph{The Journal of Chemical Physics} \textbf{2008}, \emph{129},
  244101\relax
\mciteBstWouldAddEndPuncttrue
\mciteSetBstMidEndSepPunct{\mcitedefaultmidpunct}
{\mcitedefaultendpunct}{\mcitedefaultseppunct}\relax
\EndOfBibitem
\bibitem[K{\"o}uppel \latin{et~al.}(1984)K{\"o}uppel, Domcke, and
  Cederbaum]{kouppelMultimodeMolecularDynamics1984}
K{\"o}uppel,~H.; Domcke,~W.; Cederbaum,~L.~S. In \emph{Advances in {{Chemical
  Physics}}}, 1st ed.; Prigogine,~I., Rice,~S.~A., Eds.; Wiley, 1984; Vol.~57;
  pp 59--246\relax
\mciteBstWouldAddEndPuncttrue
\mciteSetBstMidEndSepPunct{\mcitedefaultmidpunct}
{\mcitedefaultendpunct}{\mcitedefaultseppunct}\relax
\EndOfBibitem
\bibitem[K{\"o}ppel(2004)]{koppelDiabaticRepresentationMethods2004}
K{\"o}ppel,~H. \emph{Advanced {{Series}} in {{Physical Chemistry}}}; WORLD
  SCIENTIFIC, 2004; Vol.~15; pp 175--204\relax
\mciteBstWouldAddEndPuncttrue
\mciteSetBstMidEndSepPunct{\mcitedefaultmidpunct}
{\mcitedefaultendpunct}{\mcitedefaultseppunct}\relax
\EndOfBibitem
\bibitem[Gu(2023)]{guDiscreteVariableLocalDiabatic2023}
Gu,~B. A {{Discrete-Variable Local Diabatic Representation}} of {{Conical
  Intersection Dynamics}}. \emph{Journal of Chemical Theory and Computation}
  \textbf{2023}, \emph{19}, 6557--6563\relax
\mciteBstWouldAddEndPuncttrue
\mciteSetBstMidEndSepPunct{\mcitedefaultmidpunct}
{\mcitedefaultendpunct}{\mcitedefaultseppunct}\relax
\EndOfBibitem
\bibitem[Gu(2024)]{guNonadiabaticConicalIntersection2024b}
Gu,~B. Nonadiabatic {{Conical Intersection Dynamics}} in the {{Local Diabatic
  Representation}} with {{Strang Splitting}} and {{Fourier Basis}}.
  \emph{Journal of Chemical Theory and Computation} \textbf{2024}, \emph{20},
  2711--2718\relax
\mciteBstWouldAddEndPuncttrue
\mciteSetBstMidEndSepPunct{\mcitedefaultmidpunct}
{\mcitedefaultendpunct}{\mcitedefaultseppunct}\relax
\EndOfBibitem
\bibitem[Zhu and Gu(2024)Zhu, and Gu]{zhuMakingPeaceRandom2024a}
Zhu,~X.; Gu,~B. Making {{Peace}} with {{Random Phases}}: {{Ab Initio Conical
  Intersection Quantum Dynamics}} in {{Random Gauges}}. \emph{The Journal of
  Physical Chemistry Letters} \textbf{2024}, \emph{15}, 8487--8493\relax
\mciteBstWouldAddEndPuncttrue
\mciteSetBstMidEndSepPunct{\mcitedefaultmidpunct}
{\mcitedefaultendpunct}{\mcitedefaultseppunct}\relax
\EndOfBibitem
\bibitem[Xie \latin{et~al.}(2025)Xie, Liu, and
  Gu]{xieQuantumGeometricalMolecular2025}
Xie,~Y.; Liu,~R.; Gu,~B. Quantum Geometrical Molecular Dynamics. \emph{Science
  Advances} \textbf{2025}, \emph{11}, eadz3711\relax
\mciteBstWouldAddEndPuncttrue
\mciteSetBstMidEndSepPunct{\mcitedefaultmidpunct}
{\mcitedefaultendpunct}{\mcitedefaultseppunct}\relax
\EndOfBibitem
\bibitem[Sha and Gu(2026)Sha, and Gu]{shaExponentialConvergenceLocal2026}
Sha,~M.; Gu,~B. Exponential Convergence of the Local Diabatic Representation
  for Nonadiabatic Eigenvalue Problems. \emph{Physical Chemistry Chemical
  Physics} \textbf{2026}, 10.1039.D5CP03524D\relax
\mciteBstWouldAddEndPuncttrue
\mciteSetBstMidEndSepPunct{\mcitedefaultmidpunct}
{\mcitedefaultendpunct}{\mcitedefaultseppunct}\relax
\EndOfBibitem
\bibitem[Pope \latin{et~al.}(1999)Pope, Swenberg, and
  Pope]{popeElectronicProcessesOrganic1999}
Pope,~M.; Swenberg,~C.~E.; Pope,~M. \emph{Electronic Processes in Organic
  Crystals and Polymers}, 2nd ed.; Monographs on the Physics and Chemistry of
  Materials \#56; Oxford University Press: New York, 1999\relax
\mciteBstWouldAddEndPuncttrue
\mciteSetBstMidEndSepPunct{\mcitedefaultmidpunct}
{\mcitedefaultendpunct}{\mcitedefaultseppunct}\relax
\EndOfBibitem
\bibitem[Smith and Michl(2010)Smith, and Michl]{smithSingletFission2010}
Smith,~M.~B.; Michl,~J. Singlet {{Fission}}. \emph{Chemical Reviews}
  \textbf{2010}, \emph{110}, 6891--6936\relax
\mciteBstWouldAddEndPuncttrue
\mciteSetBstMidEndSepPunct{\mcitedefaultmidpunct}
{\mcitedefaultendpunct}{\mcitedefaultseppunct}\relax
\EndOfBibitem
\bibitem[Smith and Michl(2013)Smith, and Michl]{smithRecentAdvancesSinglet2013}
Smith,~M.~B.; Michl,~J. Recent {{Advances}} in {{Singlet Fission}}.
  \emph{Annual Review of Physical Chemistry} \textbf{2013}, \emph{64},
  361--386\relax
\mciteBstWouldAddEndPuncttrue
\mciteSetBstMidEndSepPunct{\mcitedefaultmidpunct}
{\mcitedefaultendpunct}{\mcitedefaultseppunct}\relax
\EndOfBibitem
\bibitem[Minami and Nakano(2012)Minami, and
  Nakano]{minamiDiradicalCharacterView2012}
Minami,~T.; Nakano,~M. Diradical {{Character View}} of {{Singlet Fission}}.
  \emph{The Journal of Physical Chemistry Letters} \textbf{2012}, \emph{3},
  145--150\relax
\mciteBstWouldAddEndPuncttrue
\mciteSetBstMidEndSepPunct{\mcitedefaultmidpunct}
{\mcitedefaultendpunct}{\mcitedefaultseppunct}\relax
\EndOfBibitem
\bibitem[Nakano \latin{et~al.}(2012)Nakano, Minami, Fukui, Kishi, Shigeta, and
  Champagne]{nakanoFullConfigurationInteraction2012}
Nakano,~M.; Minami,~T.; Fukui,~H.; Kishi,~R.; Shigeta,~Y.; Champagne,~B. Full
  Configuration Interaction Calculations of the Second Hyperpolarizabilities of
  the {{H4}} Model Compound: {{Summation-over-states}} Analysis and Interplay
  with Diradical Characters. \emph{The Journal of Chemical Physics}
  \textbf{2012}, \emph{136}, 024315\relax
\mciteBstWouldAddEndPuncttrue
\mciteSetBstMidEndSepPunct{\mcitedefaultmidpunct}
{\mcitedefaultendpunct}{\mcitedefaultseppunct}\relax
\EndOfBibitem
\bibitem[Nakano(2017)]{nakanoOpenShellCharacterBasedMolecularDesign2017}
Nakano,~M. Open-{{Shell-Character-Based Molecular Design Principles}}:
  {{Applications}} to {{Nonlinear Optics}} and {{Singlet Fission}}. \emph{The
  Chemical Record} \textbf{2017}, \emph{17}, 27--62\relax
\mciteBstWouldAddEndPuncttrue
\mciteSetBstMidEndSepPunct{\mcitedefaultmidpunct}
{\mcitedefaultendpunct}{\mcitedefaultseppunct}\relax
\EndOfBibitem
\bibitem[Thalmann \latin{et~al.}(2024)Thalmann, Ismail, Kathir, Rodrigues,
  Thoss, Mart{\'i}n~Pend{\'a}s, and Coto]{thalmannRoleRadicalCharacter2024}
Thalmann,~K.~S.; Ismail,~K.~M.; Kathir,~R.~K.; Rodrigues,~D. J.~L.; Thoss,~M.;
  Mart{\'i}n~Pend{\'a}s,~{\'A}.; Coto,~P.~B. Role of the {{Radical Character}}
  in {{Singlet Fission}}: {{An Ab Initio}} and {{Quantum Chemical Topology
  Analysis}}. \emph{The Journal of Physical Chemistry A} \textbf{2024},
  acs.jpca.4c06380\relax
\mciteBstWouldAddEndPuncttrue
\mciteSetBstMidEndSepPunct{\mcitedefaultmidpunct}
{\mcitedefaultendpunct}{\mcitedefaultseppunct}\relax
\EndOfBibitem
\bibitem[Claudino \latin{et~al.}(2023)Claudino, Peng, Kowalski, and
  Humble]{claudinoModelingSingletFission2023a}
Claudino,~D.; Peng,~B.; Kowalski,~K.; Humble,~T.~S. Modeling {{Singlet
  Fission}} on a {{Quantum Computer}}. \emph{The Journal of Physical Chemistry
  Letters} \textbf{2023}, \emph{14}, 5511--5516\relax
\mciteBstWouldAddEndPuncttrue
\mciteSetBstMidEndSepPunct{\mcitedefaultmidpunct}
{\mcitedefaultendpunct}{\mcitedefaultseppunct}\relax
\EndOfBibitem
\bibitem[Berne \latin{et~al.}(1998)Berne, Ciccotti, and
  Coker]{berneClassicalQuantumDynamics1998}
Berne,~B.~J.; Ciccotti,~G.; Coker,~D.~F. Classical and {{Quantum Dynamics}} in
  {{Condensed Phase Simulations}}. Classical and {{Quantum Dynamics}} in
  {{Condensed Phase Simulations}}. LERICI, Villa Marigola, 1998\relax
\mciteBstWouldAddEndPuncttrue
\mciteSetBstMidEndSepPunct{\mcitedefaultmidpunct}
{\mcitedefaultendpunct}{\mcitedefaultseppunct}\relax
\EndOfBibitem
\bibitem[Tully(2023)]{tullyEhrenfestDynamicsQuantum2023}
Tully,~J.~C. Ehrenfest Dynamics with Quantum Mechanical Nuclei. \emph{Chemical
  Physics Letters} \textbf{2023}, \emph{816}, 140396\relax
\mciteBstWouldAddEndPuncttrue
\mciteSetBstMidEndSepPunct{\mcitedefaultmidpunct}
{\mcitedefaultendpunct}{\mcitedefaultseppunct}\relax
\EndOfBibitem
\bibitem[Musser \latin{et~al.}(2015)Musser, Liebel, Schnedermann, Wende, Kehoe,
  Rao, and Kukura]{musserEvidenceConicalIntersection2015}
Musser,~A.~J.; Liebel,~M.; Schnedermann,~C.; Wende,~T.; Kehoe,~T.~B.; Rao,~A.;
  Kukura,~P. Evidence for Conical Intersection Dynamics Mediating Ultrafast
  Singlet Exciton Fission. \emph{Nature Physics} \textbf{2015}, \emph{11},
  352--357\relax
\mciteBstWouldAddEndPuncttrue
\mciteSetBstMidEndSepPunct{\mcitedefaultmidpunct}
{\mcitedefaultendpunct}{\mcitedefaultseppunct}\relax
\EndOfBibitem
\bibitem[Dunning(1989)]{dunningGaussianBasisSets1989b}
Dunning,~T.~H. Gaussian Basis Sets for Use in Correlated Molecular
  Calculations. {{I}}. {{The}} Atoms Boron through Neon and Hydrogen. \emph{The
  Journal of Chemical Physics} \textbf{1989}, \emph{90}, 1007--1023\relax
\mciteBstWouldAddEndPuncttrue
\mciteSetBstMidEndSepPunct{\mcitedefaultmidpunct}
{\mcitedefaultendpunct}{\mcitedefaultseppunct}\relax
\EndOfBibitem
\bibitem[Kendall \latin{et~al.}(1992)Kendall, Dunning, and
  Harrison]{kendallElectronAffinitiesFirstrow1992}
Kendall,~R.~A.; Dunning,~T.~H.; Harrison,~R.~J. Electron Affinities of the
  First-Row Atoms Revisited. {{Systematic}} Basis Sets and Wave Functions.
  \emph{The Journal of Chemical Physics} \textbf{1992}, \emph{96},
  6796--6806\relax
\mciteBstWouldAddEndPuncttrue
\mciteSetBstMidEndSepPunct{\mcitedefaultmidpunct}
{\mcitedefaultendpunct}{\mcitedefaultseppunct}\relax
\EndOfBibitem
\bibitem[Chauvin \latin{et~al.}(2016)Chauvin, Lepetit, Silvi, and
  Alikhani]{chauvinApplicationsTopologicalMethods2016}
Chauvin,~R., Lepetit,~C., Silvi,~B., Alikhani,~E., Eds. \emph{Applications of
  {{Topological Methods}} in {{Molecular Chemistry}}}; Challenges and
  {{Advances}} in {{Computational Chemistry}} and {{Physics}}; Springer
  International Publishing: Cham, 2016; Vol.~22\relax
\mciteBstWouldAddEndPuncttrue
\mciteSetBstMidEndSepPunct{\mcitedefaultmidpunct}
{\mcitedefaultendpunct}{\mcitedefaultseppunct}\relax
\EndOfBibitem
\bibitem[Mart{\'i}n~Pend{\'a}s and Francisco(2021)Mart{\'i}n~Pend{\'a}s, and
  Francisco]{martinpendasLocalSpinOpen2021}
Mart{\'i}n~Pend{\'a}s,~A.; Francisco,~E. Local Spin and Open Quantum Systems:
  Clarifying Misconceptions, Unifying Approaches. \emph{Physical Chemistry
  Chemical Physics} \textbf{2021}, \emph{23}, 8375--8392\relax
\mciteBstWouldAddEndPuncttrue
\mciteSetBstMidEndSepPunct{\mcitedefaultmidpunct}
{\mcitedefaultendpunct}{\mcitedefaultseppunct}\relax
\EndOfBibitem
\bibitem[Zirzlmeier \latin{et~al.}(2015)Zirzlmeier, Lehnherr, Coto, Chernick,
  Casillas, Basel, Thoss, Tykwinski, and
  Guldi]{zirzlmeierSingletFissionPentacene2015}
Zirzlmeier,~J.; Lehnherr,~D.; Coto,~P.~B.; Chernick,~E.~T.; Casillas,~R.;
  Basel,~B.~S.; Thoss,~M.; Tykwinski,~R.~R.; Guldi,~D.~M. Singlet Fission in
  Pentacene Dimers. \emph{Proceedings of the National Academy of Sciences}
  \textbf{2015}, \emph{112}, 5325--5330\relax
\mciteBstWouldAddEndPuncttrue
\mciteSetBstMidEndSepPunct{\mcitedefaultmidpunct}
{\mcitedefaultendpunct}{\mcitedefaultseppunct}\relax
\EndOfBibitem
\bibitem[Casanova(2018)]{casanovaTheoreticalModelingSinglet2018}
Casanova,~D. Theoretical {{Modeling}} of {{Singlet Fission}}. \emph{Chemical
  Reviews} \textbf{2018}, \emph{118}, 7164--7207\relax
\mciteBstWouldAddEndPuncttrue
\mciteSetBstMidEndSepPunct{\mcitedefaultmidpunct}
{\mcitedefaultendpunct}{\mcitedefaultseppunct}\relax
\EndOfBibitem
\bibitem[Roseiro and Robert(2022)Roseiro, and
  Robert]{roseiroEnvironmentalEffectsSinglet2022}
Roseiro,~P.; Robert,~V. Environmental Effects on the Singlet Fission
  Phenomenon: A Model {{Hamiltonian-based}} Study. \emph{Physical Chemistry
  Chemical Physics} \textbf{2022}, \emph{24}, 15945--15950\relax
\mciteBstWouldAddEndPuncttrue
\mciteSetBstMidEndSepPunct{\mcitedefaultmidpunct}
{\mcitedefaultendpunct}{\mcitedefaultseppunct}\relax
\EndOfBibitem
\bibitem[Littlejohn \latin{et~al.}(2002)Littlejohn, Cargo, Carrington,
  Mitchell, and Poirier]{littlejohnGeneralFrameworkDiscrete2002}
Littlejohn,~R.~G.; Cargo,~M.; Carrington,~T.; Mitchell,~K.~A.; Poirier,~B. A
  General Framework for Discrete Variable Representation Basis Sets. \emph{The
  Journal of Chemical Physics} \textbf{2002}, \emph{116}, 8691--8703\relax
\mciteBstWouldAddEndPuncttrue
\mciteSetBstMidEndSepPunct{\mcitedefaultmidpunct}
{\mcitedefaultendpunct}{\mcitedefaultseppunct}\relax
\EndOfBibitem
\bibitem[Light and Carrington(2000)Light, and
  Carrington]{lightDiscreteVariableRepresentationsTheir2000}
Light,~J.~C.; Carrington,~T. In \emph{Advances in {{Chemical Physics}}}, 1st
  ed.; Prigogine,~I., Rice,~S.~A., Eds.; Wiley, 2000; Vol. 114; pp
  263--310\relax
\mciteBstWouldAddEndPuncttrue
\mciteSetBstMidEndSepPunct{\mcitedefaultmidpunct}
{\mcitedefaultendpunct}{\mcitedefaultseppunct}\relax
\EndOfBibitem
\bibitem[Von~Busch \latin{et~al.}(1998)Von~Busch, Dev, Eckel, Kasahara, Wang,
  Demtr{\"o}der, Sebald, and Meyer]{vonbuschUnambiguousProofBerrys1998}
Von~Busch,~H.; Dev,~V.; Eckel,~H.-A.; Kasahara,~S.; Wang,~J.;
  Demtr{\"o}der,~W.; Sebald,~P.; Meyer,~W. Unambiguous {{Proof}} for
  {{Berry}}'s {{Phase}} in the {{Sodium Trimer}}: {{Analysis}} of the
  {{Transition A}} 2 {{E}} {$\prime$} {$\prime$} \textleftarrow{} {{X}} 2 {{E}}
  {$\prime$}. \emph{Physical Review Letters} \textbf{1998}, \emph{81},
  4584--4587\relax
\mciteBstWouldAddEndPuncttrue
\mciteSetBstMidEndSepPunct{\mcitedefaultmidpunct}
{\mcitedefaultendpunct}{\mcitedefaultseppunct}\relax
\EndOfBibitem
\bibitem[Xie \latin{et~al.}(2016)Xie, Ma, Zhu, Yarkony, Xie, and
  Guo]{xieNonadiabaticTunnelingPhotodissociation2016b}
Xie,~C.; Ma,~J.; Zhu,~X.; Yarkony,~D.~R.; Xie,~D.; Guo,~H. Nonadiabatic
  {{Tunneling}} in {{Photodissociation}} of {{Phenol}}. \emph{Journal of the
  American Chemical Society} \textbf{2016}, \emph{138}, 7828--7831\relax
\mciteBstWouldAddEndPuncttrue
\mciteSetBstMidEndSepPunct{\mcitedefaultmidpunct}
{\mcitedefaultendpunct}{\mcitedefaultseppunct}\relax
\EndOfBibitem
\bibitem[Yuan \latin{et~al.}(2018)Yuan, Guan, Chen, Zhao, Yu, Luo, Tan, Xie,
  Wang, Sun, Zhang, and Yang]{yuanObservationGeometricPhase2018}
Yuan,~D.; Guan,~Y.; Chen,~W.; Zhao,~H.; Yu,~S.; Luo,~C.; Tan,~Y.; Xie,~T.;
  Wang,~X.; Sun,~Z.; Zhang,~D.~H.; Yang,~X. Observation of the Geometric Phase
  Effect in the {{H}} + {{HD}} {$\rightarrow$} {{H}}{\textsubscript{2}} + {{D}}
  Reaction. \emph{Science} \textbf{2018}, \emph{362}, 1289--1293\relax
\mciteBstWouldAddEndPuncttrue
\mciteSetBstMidEndSepPunct{\mcitedefaultmidpunct}
{\mcitedefaultendpunct}{\mcitedefaultseppunct}\relax
\EndOfBibitem
\bibitem[Xie and Gu(2025)Xie, and Gu]{xieLinkedProductApproximation2025}
Xie,~Y.; Gu,~B. Linked {{Product Approximation}} to the {{Global Electronic
  Overlap Matrix}}. \emph{Journal of Chemical Theory and Computation}
  \textbf{2025}, \emph{21}, 9249--9258\relax
\mciteBstWouldAddEndPuncttrue
\mciteSetBstMidEndSepPunct{\mcitedefaultmidpunct}
{\mcitedefaultendpunct}{\mcitedefaultseppunct}\relax
\EndOfBibitem
\bibitem[Tully(1998)]{tullyMixedQuantumClassical1998}
Tully,~J.~C. Mixed Quantum--Classical Dynamics. \emph{Faraday Discussions}
  \textbf{1998}, \emph{110}, 407--419\relax
\mciteBstWouldAddEndPuncttrue
\mciteSetBstMidEndSepPunct{\mcitedefaultmidpunct}
{\mcitedefaultendpunct}{\mcitedefaultseppunct}\relax
\EndOfBibitem
\bibitem[Sun \latin{et~al.}(2020)Sun, Zhang, Banerjee, Bao, Barbry, Blunt,
  Bogdanov, Booth, Chen, Cui, Eriksen, Gao, Guo, Hermann, Hermes, Koh, Koval,
  Lehtola, Li, Liu, Mardirossian, McClain, Motta, Mussard, Pham, Pulkin,
  Purwanto, Robinson, Ronca, Sayfutyarova, Scheurer, Schurkus, Smith, Sun, Sun,
  Upadhyay, Wagner, Wang, White, Whitfield, Williamson, Wouters, Yang, Yu, Zhu,
  Berkelbach, Sharma, Sokolov, and Chan]{sunRecentDevelopmentsSCF2020}
Sun,~Q. \latin{et~al.}  Recent Developments in the {{P}} {\textsc{y}} {{SCF}}
  Program Package. \emph{The Journal of Chemical Physics} \textbf{2020},
  \emph{153}, 024109\relax
\mciteBstWouldAddEndPuncttrue
\mciteSetBstMidEndSepPunct{\mcitedefaultmidpunct}
{\mcitedefaultendpunct}{\mcitedefaultseppunct}\relax
\EndOfBibitem
\bibitem[Xie \latin{et~al.}(2026)Xie, Zhu, and Gu]{xiePyQEDPythonFramework2026}
Xie,~Y.; Zhu,~X.; Gu,~B. {{PyQED}}: {{A Python Framework}} for {{Ab Initio
  Geometric Quantum Dynamics}}. \emph{Chin. J. Chem. Phys.} \textbf{2026},
  \relax
\mciteBstWouldAddEndPunctfalse
\mciteSetBstMidEndSepPunct{\mcitedefaultmidpunct}
{}{\mcitedefaultseppunct}\relax
\EndOfBibitem
\end{mcitethebibliography}
	
	\newpage

	\includegraphics[width=3.25in]{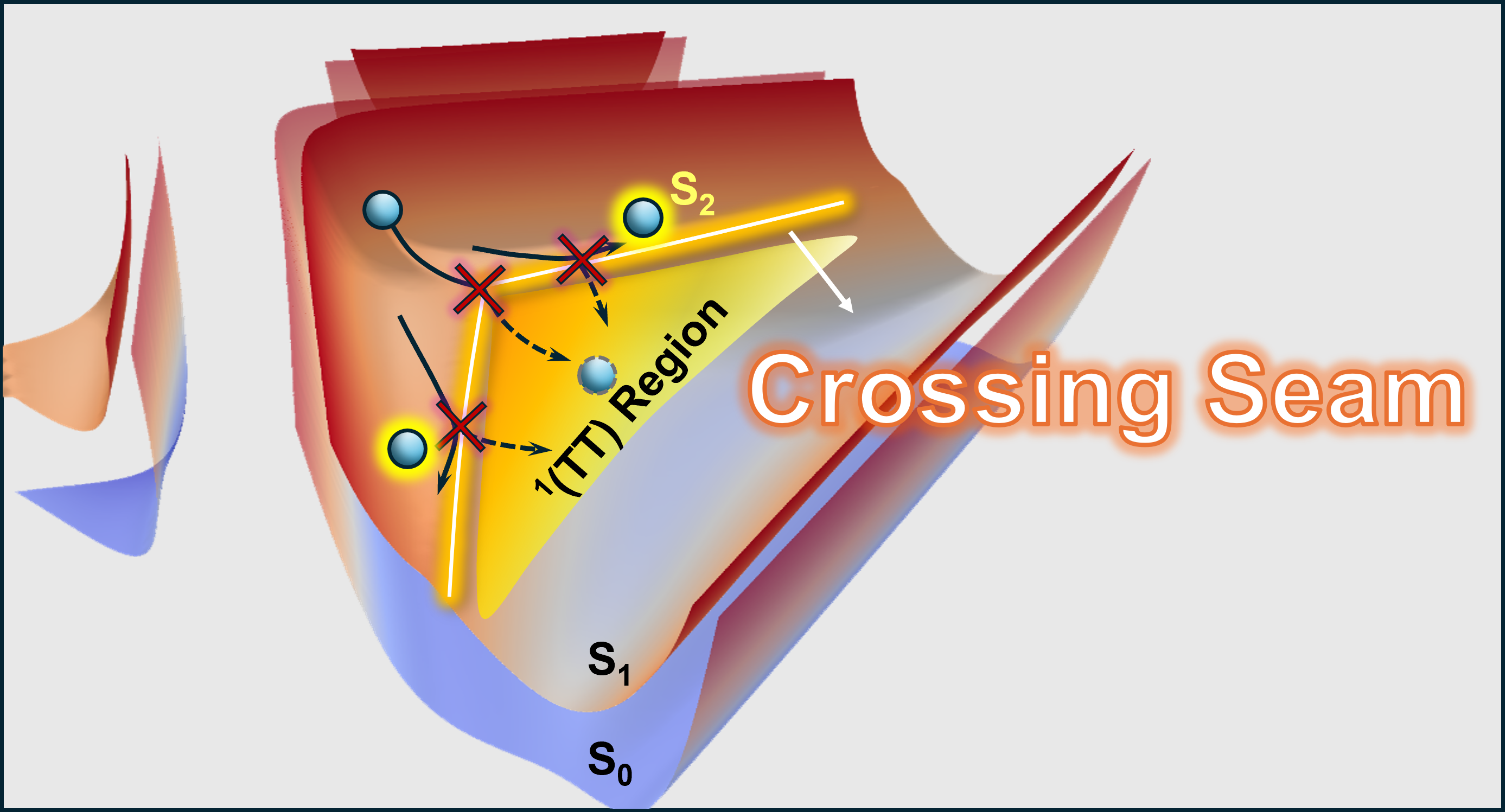}
	
\end{document}